\documentclass[10pt,journal,compsoc]{IEEEtran}
\ifCLASSOPTIONcompsoc
  \usepackage[nocompress]{cite}
\else
  \usepackage{cite}
\fi
\ifCLASSINFOpdf
\else
\fi
\usepackage{multirow}

\usepackage{amsmath}
\usepackage{amsthm}
\usepackage{amsfonts, amssymb}
\newtheorem{theorem}{Theorem}

\usepackage{ragged2e}

\usepackage{algorithmic}
\usepackage{algorithm}

\usepackage{array}
\usepackage{url}

\usepackage{graphicx}
\usepackage{bbm}
\usepackage[dvipsnames]{xcolor}

\allowdisplaybreaks[3]

\begin{document}
%
\title{Uncovering the Over-smoothing Challenge in Image Super-Resolution: Entropy-based Quantification and Contrastive Optimization}
%
%
%
%

\author{Tianshuo~Xu,
        Lijiang~Li,
        Peng~Mi,
        Xiawu~Zheng,
        Fei~Chao,~\IEEEmembership{Member,~IEEE,}
        Rongrong~Ji,~\IEEEmembership{Senior Member,~IEEE,}
        Yonghong~Tian,~\IEEEmembership{Fellow,~IEEE,}
        and~Qiang~Shen
\IEEEcompsocitemizethanks{
\IEEEcompsocthanksitem T. Xu, L. Li, and P. Mi are with the Key Laboratory of Multimedia Trusted Perception and Efficient Computing, Ministry of Education of China, School of Informatics, Xiamen University, 361005, P.R. China. 
E-mails: xutianshuo@stu.xmu.edu.cn, lilijiang@stu.xmu.edu.cn, mipeng@stu.xmu.edu.cn.
\IEEEcompsocthanksitem X. Zheng is with the Peng Cheng Laboratory, Shenzhen 518066, China, and also with the Key Laboratory of Multimedia Trusted Perception and Efficient Computing, Ministry of Education of China, School of Informatics, Xiamen University, 361005, P.R. China. E-mail: zhengxw01@pcl.ac.cn.
\IEEEcompsocthanksitem F. Chao is with the Key Laboratory of Multimedia Trusted Perception and Efficient Computing, Ministry of Education of China, School of Informatics, Xiamen University, 361005, P.R. China, and also with the Department of Computer Science, Institute of Mathematics, Physics and Computer Science, Aberystwyth University, UK. SY23 3DB. E-mail: fchao@xmu.edu.cn.
\IEEEcompsocthanksitem R. Ji is with the Key Laboratory of Multimedia Trusted Perception and Efficient Computing, Ministry of Education of China, School of Informatics, Xiamen University, 361005, P.R. China, and also with the Institute of Artificial Intelligence, Xiamen University, Xiamen 361005, China. E-mail: rrji@xmu.edu.cn.
\IEEEcompsocthanksitem Y. Tian is with the Peng Cheng Laboratory, Shenzhen 518066, China, and also with the National Engineering Laboratory for Video Technology (NELVT), School of Electronics Engineering and Computer Science, Peking University, Beijing 100871, China. E-mail: yhtian@pku.edu.cn.
\IEEEcompsocthanksitem Q. Shen is a Fellow of the United Kingdom Royal Academy of Engineering, and with the Department of Computer Science, Institute of Mathematics, Physics and Computer Science, Aberystwyth University, UK. SY23 3DB. Email: qqs@aber.ac.uk.
}
\thanks{(Corresponding author: Fei Chao.)}
\thanks{(Author T. Xu and L. Li contributed equally to this research.)}
\thanks{Manuscript revised December 10, 2023.}
}

%
%

\markboth{IEEE TRANSACTIONS ON PATTERN ANALYSIS AND MACHINE INTELLIGENCE, VOL. XX, NO. XX}{}%

%



\IEEEtitleabstractindextext{%
\begin{abstract}
\justifying PSNR-oriented models are a critical class of super-resolution models with applications across various fields. However, these models tend to generate over-smoothed images, a problem that has been analyzed previously from the perspectives of models or loss functions, but without taking into account the impact of data properties. In this paper, we present a novel phenomenon that we term the center-oriented optimization (COO) problem, where a model's output converges towards the center point of similar high-resolution images, rather than towards the ground truth. We demonstrate that the strength of this problem is related to the uncertainty of data, which we quantify using entropy. We prove that as the entropy of high-resolution images increases, their center point will move further away from the clean image distribution, and the model will generate over-smoothed images. Implicitly optimizing the COO problem, perceptual-driven approaches such as perceptual loss, model structure optimization, or GAN-based methods can be viewed. We propose an explicit solution to the COO problem, called Detail Enhanced Contrastive Loss (DECLoss). DECLoss utilizes the clustering property of contrastive learning to directly reduce the variance of the potential high-resolution distribution and thereby decrease the entropy. We evaluate DECLoss on multiple super-resolution benchmarks and demonstrate that it improves the perceptual quality of PSNR-oriented models. Moreover, when applied to GAN-based methods, such as RaGAN, DECLoss helps to achieve state-of-the-art performance, such as 0.093 LPIPS with 24.51 PSNR on 4× downsampled Urban100, validating the effectiveness and generalization of our approach.
\end{abstract}
}

\maketitle

\IEEEdisplaynontitleabstractindextext

%
\IEEEpeerreviewmaketitle

\IEEEraisesectionheading{\section{Introduction}\label{sec:introduction}}

%
%
%
%
\IEEEPARstart{I}{mage} super-resolution (SR) is a crucial task in image processing, which involves constructing a high-resolution (HR) image from its corresponding low-resolution (LR) version. 
\begin{figure}[!t]
\centering
\includegraphics[width=\linewidth]{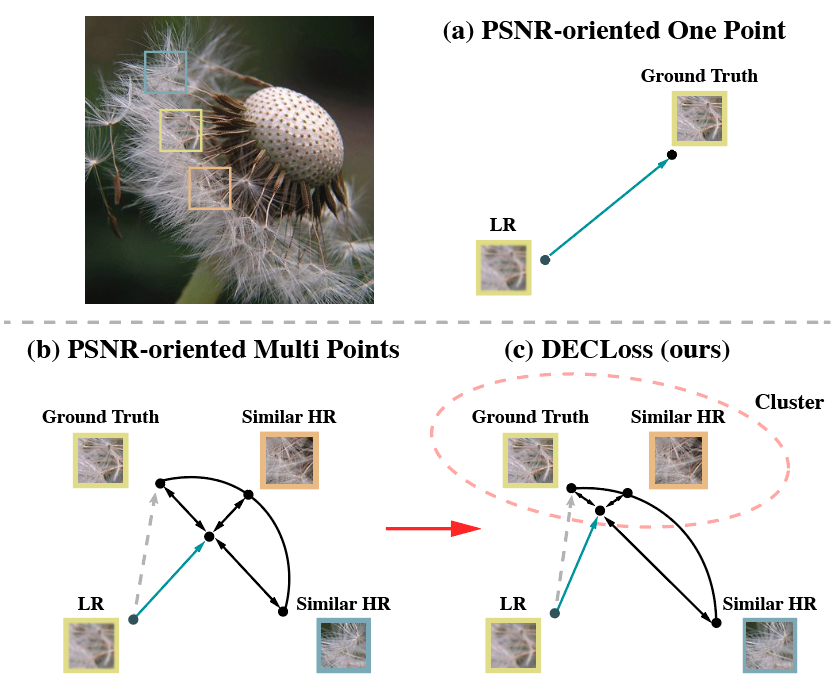}
\caption{The schematic diagram of the COO problem. (a) shows the ideal mapping from low-resolution (LR) to high-resolution (HR). As the number of similar HRs increases, (b) illustrates the actual mapping under the PSNR-oriented methods. The model tends to produce over-smoothed HR images that converge to the center point of similar HRs rather than the ground truth. To address this issue, we introduce the Detail Enhanced Contrastive Loss (DECLoss), depicted in (c), which employs the cluster property of contrastive learning to reduce the influence of the COO problem.}
\label{fig:coo_node}
\end{figure}
It has found wide-ranging applications in real-world scenarios, such as medical imaging \cite{sr_medical,sr_medical_3d}, surveillance \cite{sr_surveillance}, and face recognition \cite{sr_face}. In recent years, deep learning models have driven rapid progress in SR, including SRCNN \cite{srcnn}, VDSR \cite{vdsr}, SRResNet \cite{srresnet}, EDSR \cite{edsr}, and ESRGAN \cite{esrgan}. These models typically rely on a pixel-level mapping from lower resolutions to higher resolutions and use loss functions such as mean average error (MAE) or mean square error (MSE) to optimize for high Peak Signal to Noise Ratio (PSNR), thus dubbed PSNR-oriented models.

However, PSNR-oriented models often encounter the problem of generating over-smoothed images that lack fine details and textures \cite{srresnet, perceptual_loss, esrgan}. Despite many efforts to enhance the perceptual quality of SR images or analyze the model and loss function level \cite{pd}\cite{pd1}, the over-smoothed problem still requires a rigorous theoretical analysis and formulation based on the property of data, \textit{e.g.,} the variation of the data distribution and the amount of data. In this paper, we present the first attempt to characterize this problem from the perspective of analytic geometry and linear algebra. We propose two theorems revealing that under the constraints of existing Euclidean distance-based loss functions, PSNR-oriented models tend to generate pixels that do not converge to the ground truth HR pixels. Rather, they converge to the midpoint of many similar HR pixels (or the center of potential HR distribution). We term this phenomenon as Center-oriented Optimization (COO) Problem.


To better intuitively understand the COO problem, we consider two different HR images (or patches) that are downsampled to two very similar LR images (or patches). The model cannot distinguish them and treats them as one LR image.
In this case, the generated image under the constraint of the Euclidean distance loss function will tend to be closer to the midpoint of the two HR images and away from both HR images. SRGAN \cite{srresnet} was the pioneering work, pointed out that MSE-based methods tend to produce over-smoothed images by averaging possible high-resolution candidates. Extending this analysis, we discover that as the number of HR images increases, the generated image will gradually converge to a fixed point (Fig.~\ref{fig:coo_node} (b)). Consider an interpretation from the perspective of analytical geometry: all HR images (corresponding to a given LR image) can be regarded as a hyperplane, and this fixed point is the blurriest point because this point is the closest point to the given LR in the hyperplane. 

To further investigate the COO problem, in this paper, we utilized analytical geometry to analyze and propose two theorems to describe how data uncertainty affects the quality of generated images. We introduce the concept of information entropy \cite{kullback1997information} to measure the uncertainty of the data, and we prove that as the entropy increases, the generated image (subject to a Euclidean distance constraint) will converge to the blurriest point on the hyperplane. We also reveal that the distance between the generated image and the real HR images will increase with higher data entropy, implying that more uncertain data leads to less realistic image generation. The theory of the COO problem is formed by these two theorems that demonstrate the effect of entropy on the quality of generated images. This theory extends the current research on image distortion and provides an alternative framework for addressing over-smoothing problems.

Existing methods implicitly address this problem from two main aspects. The first one is to enhance the model’s learning ability by optimizing the network structure or adding more parameters, such as SwinIR \cite{swin} and RRDB \cite{esrgan}. As the learning ability improves, the model can better fit the joint distributions of LRs and HRs, thereby reducing the entropy of potential HRs. The second perspective is to model HR distributions, instead of minimizing the Euclidean distance between generated images and ground truth HRs, that is, transforming SR from a regression problem into a conditional generation problem. In practice, some models add GAN-based losses to optimize the PSNR-oriented models \cite{srresnet, esrgan, ranksrgan, pan2020physics} or directly build a generative framework for SR, for example, flow-based model \cite{srflow} or diffusion-based model \cite{sr3, chen2016trainable}. However, both perspectives have obvious drawbacks. Generally, enhancing learning ability requires numerous experimental trials, which are often costly. Meanwhile, applying a generative network to SR will make the generated results more different from the original LR \cite{wang2023gan}.


In this paper, we present a more explicit solution to the COO problem, which is to directly reduce the variance of the potential HR distribution. To this end, we devise the \emph{Detail Enhanced Contrastive Loss} (DECLoss), which exploits the clustering property of contrastive learning to address this problem, shown in Fig.~\ref{fig:coo_node} (c). Specifically, we represent the potential HR patch distribution (of a given LR patch) as a Gaussian distribution, with the ground truth HR patch as the mean value. Then, we identify the HR patches within a finite range of distances from the mean value as positive samples and the others as negative samples, and optimize them with a contrastive learning approach \cite{simclr} to minimize the distance between positive samples and maximize the distance between negative samples. Furthermore, by utilizing the clustering property of contrastive learning, we can reduce the variance of this distribution by increasing the probabilities of a few HR pixels near the mean value and decreasing the probabilities of others, thus reducing the information entropy and enabling the model to generate more visually pleasing results.

To evaluate the effectiveness of the proposed DECLoss, we perform comprehensive experiments on multiple SR benchmarks. The experimental results demonstrate that our DECLoss can improve the perceptual quality of the PSNR-oriented models, such as RRDB \cite{esrgan}, and it is also compatible with GAN-based methods like RaGAN \cite{esrgan}. More importantly, with the assistance of DECLoss, these methods can surpass a variety of state-of-the-art methods on perceptual metric LPIPS \cite{lpips}, while preserving a high PSNR score. For example, from 4x bicubic downsampling, DECLoss achieves 0.093 LPIPS with 24.51 PSNR on Urban100 \cite{urban100} and 0.080 LPIPS with 28.19 PSNR on the DIV2K validation set \cite{div2k}.

In summary, our main contributions are three-fold:
\begin{itemize}
  \item We identify and define the center-oriented optimization (COO) problem, which is a critical issue that causes PSNR-oriented methods to generate over-smoothed images. Our novel approach examined this problem from the perspective of analytic geometry, revealing it for the first time in this context.
  \item We established two theorems that explain how data entropy affects the COO problem and the perceptual quality of generated images. These theorems provide a new framework for optimizing over-smoothed problems, and reveal that many existing methods implicitly alleviate the COO problem from different perspectives.
  \item We devise an explicit solution called Detail Enhanced Contrastive Loss (DECLoss) that uses contrastive learning to reduce the entropy of the potential HR distribution. Extensive experiments demonstrate that DECLoss enhances both PSNR-oriented and GAN-based models in terms of perceptual quality and achieves state-of-the-art performance.
\end{itemize}


The remainder of this paper is organized as follows. In Sec.~\ref{sec:related}, we review the related work on image super-resolution, contrastive learning, and the range-null space decomposition. In Sec.~\ref{sec:coo_problem}, we reveal the center-oriented optimization (COO) problem in detail and provide the corresponding theory and proof. We present our proposed method in Sec.~\ref{sec:method}, including the loss function and the training procedure. In Sec.~\ref{sec:experiment}, we report the experimental results on several benchmark datasets and compare our method with state-of-the-art algorithms. Finally, we conclude the paper and discuss future work in Sec.~\ref{sec:conclu}.


\section{Related Work}\label{sec:related}
\subsection{Image Super-Resolution}
Image super-resolution is a significant image restoration task in computer vision, with a rich history of research that has led to notable advancements in the field. Early works such as SRCNN \cite{srcnn} utilized early convolutional neural networks to tackle SR tasks. VDSR \cite{vdsr} followed this trend by using a very deep network to improve the results. ResNet, introduced by He \emph{et al.} \cite{resnet} for learning residuals, inspired the development of SRResNet \cite{srresnet} that integrated ResBlock \cite{resnet} to enhance network depth. Building on this foundation, EDSR \cite{edsr} further improved residual methods for more efficient super-resolution results. Other methods, such as DRCN \cite{drcn}, DRRN \cite{drrn}, and CARN \cite{carn}, also incorporated ResBlock \cite{resnet} for recursive learning. Later advancements, such as RRDB \cite{esrgan} and RDN \cite{rdn}, leveraged dense connections to facilitate the transfer of information from previous layers. Attention mechanisms were also introduced to deep super-resolution models, such as RCAN \cite{rcan} and RFA \cite{rfa}, to focus the network's attention on relevant features. These advances have helped to create a diverse range of methods and models that have made significant contributions to the field of image super-resolution.

To enhance the perceptual quality and capture missing details, several approaches have been proposed in the literature. One such approach is perceptual loss, which was introduced by Johnson \emph{et al.} \cite{perceptual_loss}. Another method, called LPIPS, was proposed by Zhang \emph{et al.} \cite{lpips} to measure the perceptual distances between images. The SRGAN \cite{srresnet} model, on the other hand, incorporated a GAN-based method into an SR model, leveraging an adversarial loss during training. This paved the way for the ESRGAN \cite{esrgan} model, which enhanced the adversarial loss function and backbone structure beyond the SRGAN framework. These modifications collectively contributed to the generation of more realistic images.

Another line of research involves transforming SR into a generative framework. For instance, the SRFlow \cite{srflow} and SR3 \cite{sr3} models utilized flow-based and DDPM\cite{ho2020ddpm}-based frameworks, respectively. Although perceptual-oriented methods have proven to be effective in generating fake details, the resulting images often differ significantly from the ground truth. Therefore, there is still much research to be done in this area to achieve more accurate and realistic super-resolution.


\subsection{Correlation Analysis between Perceptual Quality Metrics and PSNR-oriented Methods}
Yochai \textit{et al.} \cite{pd} proposed that there exists a trade-off between perceptual quality and distortion measures for all image restoration tasks. The distortion measures refer to the dissimilarity between the reconstructed image and the ground truth, such as peak signal-to-noise ratio (PSNR) and structural similarity index (SSIM)\cite{ssim}. The perceptual quality of a reconstructed image is the degree to which it looks like a natural image. The perceptual-distortion trade-off demonstrates that there exists a region in the perception-distortion plane that cannot be attained, \emph{i.e.}, an image restoration algorithm can only improve either distortion or perceptual quality, but not both. Many studies focus on improving the image restoration algorithm to approach the boundary of the perception-distortion plane \cite{pd,pd1}. For example, ref.~\cite{pd} proposed to utilize GANs to achieve a better trade-off between distortion and perceptual quality. The theory about perceptual-distortion trade-off explains why optimizing PSNR (\emph{i.e.,} low distortion) leads to blurry (\emph{i.e.,} low perceptual quality) images from an algorithm perspective, while we explain the reason from a \textit{data perspective} in this paper. Specifically, we prove that the entropy of data is directly connected to the perceptual quality of images generated by PSNR-oriented algorithms. Given a fixed image restoration algorithm, the algorithm tends to generate blurrier images as the entropy increases. We believe that this conclusion has guiding significance for the model and loss function design in future research.

\subsection{Contrastive Learning}
Contrastive Learning has demonstrated its effectiveness in unsupervised representation learning tasks \cite{byol, contrast_representation, contrast_understanding}. The objective of a contrastive learning method is to learn representations that maintain the proximity between similar samples while keeping dissimilar samples far apart \cite{simclr, moco, caron2020unsupervised}. Despite achieving remarkable success on various high-level tasks, there are fewer contrastive learning methods proposed for low-level tasks such as super-resolution \cite{wu2021practical, wang2021unsupervised, zhang2021blind}. Vanilla contrastive learning technologies developed for high-level visual tasks are not directly applicable to low-level visual tasks, as they generate global visual representations that are insufficient for tasks requiring rich texture and context information. Recent works have explored various ways of applying contrastive learning to super-resolution problems. For instance, Zhang \emph{et al.} \cite{zhang2021blind} proposed a contrastive representation learning network that focuses on the blind super-resolution of images with multi-modal and spatially variant distributions. They introduced contrastive learning to extract resolution-invariant embedding and discard resolution-variant embedding under the guidance of a bidirectional contrastive loss. Wu \emph{et al.} \cite{wu2021practical} developed a practical contrastive learning framework for single image super-resolution by generating numerous informative positive and hard negative samples in the frequency space. DASR \cite{wang2021unsupervised} applied contrastive learning in the degradation representations, and they also designed a simple but effective embedding network inherited from the discriminator network. However, these studies primarily focused on the comparison learning algorithm itself. In this paper, inspired by our proposed theorems, we utilize contrastive learning to constrain the information entropy, thereby improving the perceptual quality of the super-resolved images.

\subsection{Range-Null Space Decomposition}
The Range-Null decomposition (RND) is a fundamental concept in linear algebra that relates to any linear operator or matrix $\mathbf{A}$. It states that the domain of $\mathbf{A}$ can be partitioned into two orthogonal subspaces: the null space $\mathcal{N}$($\mathbf{A}$) and the range space $\mathcal{R}$($\mathbf{A}$). The null space $\mathcal{N}$($\mathbf{A}$) consists of all vectors $\mathbf{x}$ that satisfy $\mathbf{Ax}=0$, which is also called the kernel of the operator or matrix. Conversely, the range space $\mathcal{R}$($\mathbf{A}$) is the set of all vectors that can be mapped by the operator or matrix and is also termed the image of the operator or matrix. The Range-Null decomposition is a powerful mathematical tool that has numerous applications, such as solving image inverse problems like image restoration, super-resolution, and deblurring. Wang \textit{et al.} \cite{wang2021towards} recently applied null-space analysis in Generative Adversarial Networks (GANs) and modeled the zero-domain using GAN prior. DDNM \cite{wang2022zero}, a zero-shot image restoration method, leverages a diffusion model to generate realistic null-space contents that satisfy data consistency. Our proposed theorems are relevant to the Range-Null analysis, which we will discuss in detail in Sec.~\ref{sec:rnd}.

\begin{figure*}[t]
  \centering
  \includegraphics[width=1.0\linewidth]{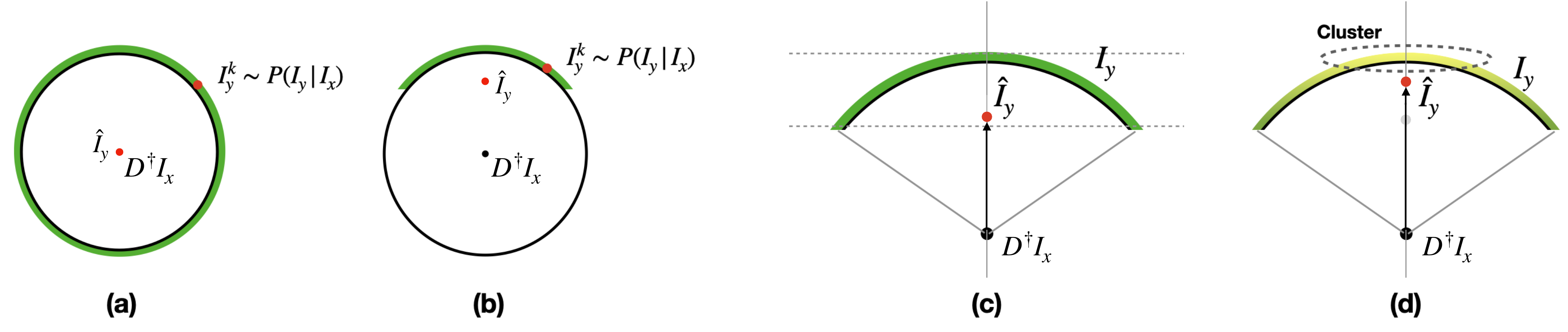}
  \caption{Illustration of the Center-Oriented Optimization (COO) problem from the geometry perspective. The green arc represents the distribution of potential high-resolution (HR) images $I_y$ corresponding to a given low-resolution (LR) image $I_x$, $D^\dagger I_x$ is the projection point of LR $I_x$ onto HR hyperplane, $\hat{I_y}$ denotes the mean of the HR distribution. Panel (a) shows an extreme case where the entropy of the HR distribution reaches the upper bound of the space. Panel (b) is a general case, each LR pixel has a finite number of potential HR pixels, which is further detailed in panel (c). Panel (d) describes the principle of our proposed method DECLoss, which reduces the variance of the HR distribution using contrastive learning to alleviate the COO problem.}
  \label{fig:coo_problem}
\end{figure*}

\section{Center-oriented Optimization Problem}\label{sec:coo_problem}
The basic concept of the Center-oriented Optimization Problem (COO) is: if two very similar LRs as input, an SR model will regard the two similar LRs as one image, but their corresponding HRs are different from each other. Under the constraint of Euclidean distance, the output of the model is forced to converge to the mean value between HRs' pixels, i.e.,  the output will be far from the HR distribution and thus appear over-smoothed. Similar issues have been analyzed in several studies \cite{pd,pd1}.

We not only provide a detailed explanation of the cause of the COO problem from the perspectives of geometry and probability but also propose a framework based on entropy and the midpoint of Euclidean distance. We mathematically prove that, under the condition of model invariance, as data entropy increases, the midpoint of Euclidean distance tends to approach the most ambiguous point. Additionally, many existing methods for improving model perceptual quality can also be explained using our framework.
\subsection{Preliminaries}
Following the symbol system theory \cite{wang2020deep}, we first define the degradation function, which obtains an LR image $I_x$ by downsampling from a corresponding HR image $I_y$,
\begin{equation}
   I_x = D(I_y; s),
\end{equation}
where $D$ denotes the degradation function and $s$ indicates a scaling factor. Image super-resolution aims at learning a function $F$ to recover HR $\hat{I_y}$ from a given LR $I_x$, 
\begin{equation}
   \hat{I_y} = F(I_x; \theta),
\end{equation}
where $\theta$ is the learnable parameter of $F$.  The optimization object of super-resolution is then defined as
\begin{equation}\label{eq:model_argmin}
   \hat{\theta} = \mathop{\arg\min}_\theta L(\hat{I_y}, I_y),
\end{equation}
where $L(\hat{I_y}, I_y)$ is the loss function between the generated image $\hat{I_y}$ and the ground truth image $I_y$. In most PSNR-oriented methods, $L$ is a Euclidean distance-based loss function, \emph{e.g.}, mean square error. 

In general, the degradation process $D$ or scaling factor $s$ is unknown. To better explain, we draw on the completeness of linear space and assume degradation $D$ is a linear matrix, then we have
\begin{equation}
    I_x = D I_y.
\end{equation}


\subsection{Problem Definition and Description}
Firstly, we focus on two concepts relevant to HR $I_y$, the HR space and the conditional HR data distribution. From the perspective of geometry, HR space is a hyperplane $I_y$, satisfying $DI_y-I_x=0$, where $I_y \in \mathbb{R}^n, I_x \in \mathbb{R}^m, D \in \mathbb{R}^{m\times n}$. Notably, $ m < n$ is permanent. Conditional HR data distribution $P(I_y|I_x)$ is the intersecting points between the real data distribution $P(I_y)$ and the hyperplane corresponds to the given $I_x$. For simplicity, we assume that $P(I_y|I_x)$ is a Gaussian distribution in the hyperplane, although the data are sampled from the real dataset,
\begin{equation}
    \mathcal{N}(I_y|\mu, \Sigma) = \frac{1}{Z}\exp \left(-\frac{1}{2}(I_y-\mu)^T \Sigma^{-1}(I_y-\mu) \right),
\end{equation}
where $Z=(2\pi)^{n/2}|\Sigma|^{1/2}$, $\Sigma$ is the co-variance matrix. 
Notably, our method's choice to model HR patch distribution as an unimodal Gaussian distribution is rooted in the large amount of data and the statistical principles of large numbers. Using a multimodal distribution is theoretically complex and hard to implement in our proposed theory, we give details in Appendix C.
Additionally, $\mu$ is the mean of $P(I_y|I_x)$ and equals the projection point $D^\dagger I_x$ of $I_x$ onto hyperplane $DI_y-I_x=0$,

\begin{equation}
    \mu = D^\dagger I_x.
\end{equation}
In this case, $D^\dagger$ is the pseudo-inverse matrix of $D$, which can be obtained from Singular Value Decomposition (SVD). Intuitively, since $D^\dagger I_x$ is the closed point to $I_x$, $D^\dagger I_x$ is the \textit{blurriest} point in the hyperplane. Then every $I_y$ in this hyperplane can be written as 
\begin{equation}\label{eq:vec_v}
    I_y = D^\dagger I_x + \vec{v},\ D\vec{v}=0.
\end{equation}
Note that, due to the assumption that $P(I_y|I_x)$ is a Gaussian distribution with $\mu$ as the mean, $I_y\sim \mathcal{N}(D^\dagger I_x, \Sigma)$, vector $\vec{v}\sim \mathcal{N}(0, \Sigma)$.

\subsection{Derivations of Information Entropy}\label{sec:entropy}
Intuitively, The divergence of the distribution and the number of points should be the two main factors affecting the Euclidean distance. We thus introduce the Information entropy to measure the uncertainty or randomness of the data distribution, the entropy of $P(I_y|I_x)$ can be defined as
\begin{equation}\label{eq:entropy}
    H(P(I_y|I_x)) = - \sum_{I_y} P(I_y|I_x) \log P(I_y|I_x).
\end{equation}

We then discuss in detail the effect of the values of $P$ and $N$ on entropy. In information entropy, assuming the amount of data ($N$) is constant, when $P=P(I_y^1)=P(I_y^2)=\cdots=P(I_y^N)=\frac{1}{N}$, $H(P(I_y|I_x))$ is taken to the point of maximum value, proved: 

Let $P(I_y^1)=p, P(I_y^i)=1-p-c_i$, where $c_i = P(I_y^2)+P(I_y^3)+\cdots+P(I_y^{i-1})+P(I_y^{i+1})+\cdots+P(I_y^N)$. Then,
\begin{align}
    f(p) &= -[P(I_y^1)\log P(I_y^1)+P(I_y^i) \log P(I_y^i)] \notag \\
    &= -[p\log p+(1-p-c_i)\log (1-p-c_i)].
\end{align}
Taking partial derivatives of $f$
\begin{equation}
    \frac{\partial f}{\partial p} = \log (1-p-c_i) - \log p.
\end{equation}
When 
\begin{equation}
    \frac{\partial f}{\partial p} = 0,
\end{equation}
\begin{equation}
    p = 1-p-c_i.
\end{equation}
Also when $p < 1-p-c_i,\ \frac{\partial f}{\partial p} > 0$, and $p > 1-p-c_i,\ \frac{\partial f}{\partial p} < 0$, therefore, when $p = 1-p-c_i$, \textit{i.e.,} when $P(I_y^1)=P(I_y^i)$, $f(p)$ is taken to the point of maximum value. Thus, when $P=P(I_y^1)=P(I_y^2)=\cdots=P(I_y^N)=\frac{1}{N}$, $H(P(I_y|I_x))$ is taken to the point of maximum value. End of the proof.

As for $N$, when $H(P(I_y|I_x))$ is taken to the point of maximum value, $P=1/N$, the formulation of $H(P(I_y|I_x))$ can be simplified as
\begin{align}\label{eq:entropy-N-start}
    H(P) & = - \sum_{I_y} P(I_y|I_x) \log P(I_y|I_x) \notag \\
    & = - \sum_{i=1}^{N} \frac{1}{N} \log \frac{1}{N} \notag \\
    & = - \log \frac{1}{N} \notag \\
    & = \log N.
\end{align}
Obviously, entropy $H(P(I_y|I_x))$ is a monotonic function that increases with the amount of data $N$.

To summarize, the information entropy $H(P(I_y|I_x))$ quantifies the uncertainty or unpredictability of the information content, which is a function of two key factors: the quantity of data and the variation of the probability distribution. The entropy is positively correlated with these factors, meaning that more data or more diverse probabilities lead to higher entropy.

\subsection{Theorems for COO Problem Description}
In this subsection, we present a detailed derivation and analysis of the relationship between information entropy and the midpoint of Euclidean distance of the HR distribution (this concept can also be viewed as the model’s output, proved in Eq.~\ref{eq:model_argmin}). We propose two fundamental theorems to characterize the COO problem. The first theorem shows that the model’s output approaches the blurriest points in the HR hyperplane as entropy increases. The second theorem demonstrates that the model’s output deviates further from the mean of the real HR distribution as entropy increases.
\begin{theorem}{(The Entropy-Midpoint Convergence)}\label{the:entropy-midpoint} 
    Let $\hat{I_y}$ denote the midpoint of the Euclidean distance of the high-resolution distribution, which has the shortest Euclidean distance to all points $(I_y^i)$ in the high-resolution distribution $P(I_y|I_x)$, and can also be regarded as the model's output (Eq.~\ref{eq:model_argmin}),
    \begin{equation}
        \hat{I_y} =\mathop{\arg\min}_{I_y^\prime} \sum_{i=1}^N d(I_y^i, I_y^\prime).
    \end{equation}
    As the entropy of a dataset increases, $\hat{I_y}$ tends to approach the point of maximum ambiguity within the hyperplane. This convergence occurs at the upper bound of the space when the entropy reaches its maximum value.
\end{theorem}

\begin{proof}
    Let $d(\cdot)$ denote the Euclidean distance, since the midpoint of the Euclidean distance of a distribution can be regarded as the distribution's expectation, we have
    \begin{align}
        \hat{I_y}&=\mathop{\arg\min}_{I_y^\prime} \sum_{i=1}^N d(I_y^i, I_y^\prime) \notag \\
        &=E(P(I_y|I_x)) \notag \\
        &=\sum_{i=1}^N p(I_y^i|I_x) I_y^i \ \ \ (\text{where} \sum_{i=1}^N p(I_y^i|I_x) = 1) \notag \\
        &=\sum_{i=1}^N p(I_y^i|I_x) (D^\dagger I_x + \vec{v}_i) \notag \\
        &=D^\dagger I_x + \sum_{i=1}^N p(I_y^i|I_x) \vec{v}_i  \notag \\
        & (\text{because}\ p(I_y^i|I_x) = p(\vec{v}_i)) \notag \notag \\
        &=D^\dagger I_x + \sum_{i=1}^N p(\vec{v}_i) \vec{v}_i.
    \end{align}

    We then calculate the distance between $\hat{I_y}$ and $D^\dagger I_x$,
    \begin{align}
        L &= |\hat{I_y} - D^\dagger I_x| \notag \\
        &= |D^\dagger I_x + \sum_{i=1}^N p(\vec{v}_i) \vec{v}_i - D^\dagger I_x| \notag \\
        &= |\sum_{i=1}^N p(\vec{v}_i) \vec{v}_i|.
    \end{align}
    After deduction, we can find that the distance between $\hat{I_y}$ and $D^\dagger I_x$ is only related to the mean value of the vector $\vec{v}$.

    Subsequently, we further derive the entropy of $P(I_y|I_x)$. From the definition of entropy (Sec.~\ref{sec:entropy}), entropy is related to the probability of each point and the number of points.

    In practice, Gaussian distribution is mainly studying the case where there are a large number of points \cite{gaussian-distribution}. However, when the sample size is small, relying solely on the Gaussian distribution to estimate entropy can lead to significant errors \cite{altman2005standard}. Therefore, we propose a rigorous method, constructing a segment function to describe the situation under different data volumes. We use a constant $\Theta > 0$ as a threshold value to determine the type of distribution. If the number of data exceeds the threshold, we assume that the data follows a Gaussian distribution. Otherwise, we adopt a uniform distribution to describe the data distribution.

    Specifically, if the number of data points $N<\Theta$, each point can be considered with equal probability ($p(\vec{v}_1)=p(\vec{v}_2)=\cdots=p(\vec{v}_N)=\frac{1}{N}$), then 
    \begin{equation}
        L = |\frac{1}{N}\sum_{i=1}^N \vec{v}_i| = \frac{1}{N}|\sum_{i=1}^N \vec{v}_i|.
    \end{equation}
    Under this consideration, the expression for entropy can be simplified as
    \begin{equation}
        H(P) = \log N\ \text{(proved in Eq.~\ref{eq:entropy-N-start})}.
    \end{equation}

    On the other hand, when the number of data points exceeds the threshold value (i.e., $N > \Theta$), 
    \begin{equation}
        L = |\sum_{i=1}^N p(\vec{v}_i) \vec{v}_i| \to E(\vec{v}).
    \end{equation}
    The expression for entropy can be computed by using the Gaussian distribution as
    \begin{figure}[!t]
    \centering
    \includegraphics[width=\linewidth]{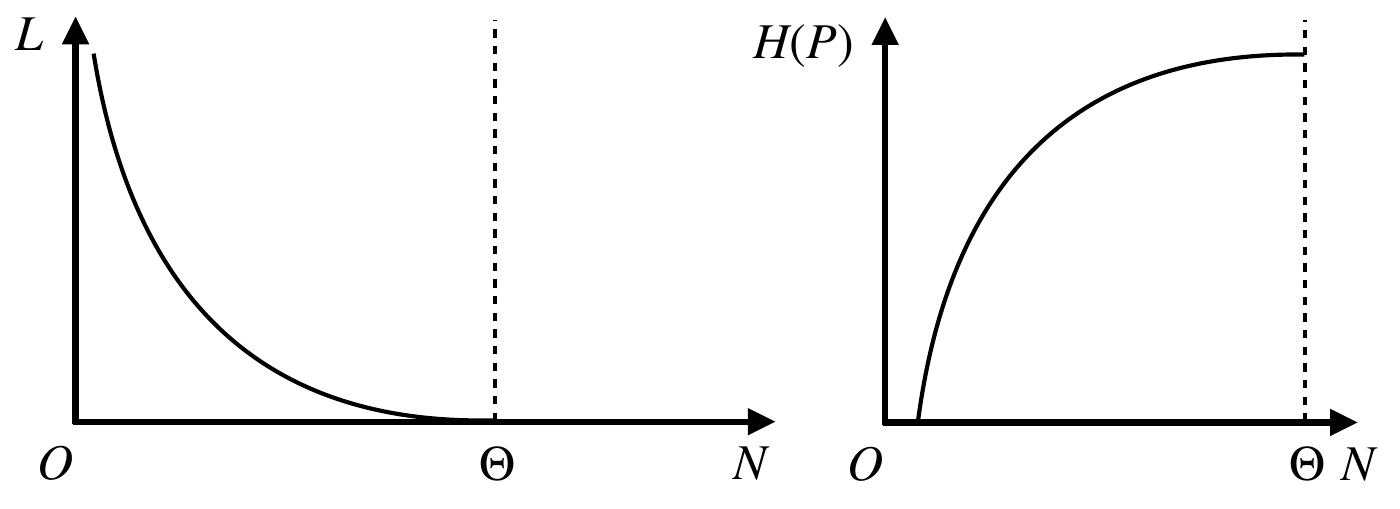}
    \caption{An illustration of Theorem~\ref{the:entropy-midpoint}, which states that the entropy $H(P)$ is positive correlated with the number of samples $N$ when $N<\Theta$. The figure also shows that the distance ($L$) between the model output and the blurriest point, decreases monotonically as $N$ increases, and reaches a lower bound when $N \geq \Theta$.
    }
    \label{fig:t1_illustrate}
    \end{figure}

    \begin{align} \label{eq:gd-entropy}
        H(P) & = - \sum_{I_y} P(I_y|I_x) \log P(I_y|I_x) \notag \\
        &= -\sum_{I_y} P(I_y|I_x) \left(-\frac{1}{2}(I_y-\mu)^T\Sigma^{-1}(I_y-\mu) - \log Z \right) \notag \\
        &= -\sum_{I_y} P(I_y|I_x) \left(-\frac{1}{2}(I_y-\mu)^T\Sigma^{-1}(I_y-\mu) \right) + \notag \\
        &\ \ \ \ \log Z \left(\sum_{I_y} P(I_y|I_x)\right) \notag \\
        &= -\sum_{I_y} P(I_y|I_x) \left(-\frac{1}{2}(I_y-\mu)^T\Sigma^{-1}(I_y-\mu) \right) + \log Z \notag \\
        &=  \sum_{I_y} P(I_y|I_x) \left( \frac{1}{2}(I_y - \mu)^T\Sigma^{-\frac{1}{2}} \cdot \Sigma^{-\frac{1}{2}} (I_y - \mu)  \right) \notag\\ &+ \log Z \notag \\
        &(\text{let } \boldsymbol{t}=\Sigma^{-\frac{1}{2}}(I_y - \mu)) \notag \\
        &=  \frac{1}{2} \sum_{I_y} P(I_y|I_x) \boldsymbol{t}^T \boldsymbol{t} +\log Z \notag \\
        &= \frac{1}{2} E \left [\boldsymbol{t}^T\boldsymbol{t}\right ] +\log Z \notag 
        \\
        &= \frac{1}{2} E \left[\text{tr}(\boldsymbol{t}\boldsymbol{t}^T)\right ] + \log Z \notag
        \\
        &= \frac{1}{2} \text{tr}(E[\boldsymbol{t}\boldsymbol{t}^T]) + \log Z \notag
        \\
        &= \frac{1}{2} \text{tr}(\mathbb{I}) + \log Z  \ \ \ (\text{where $\mathbb{I}$ is an unit matrix} )\notag \\
        &= \frac{n}{2} + \log Z \notag \\
        &= \frac{n}{2} + \log \left((2\pi)^{\frac{n}{2}} \cdot |\Sigma|^{\frac{1}{2}}\right) \notag \\
        &=\frac{n}{2}(\log 2\pi + 1) + \frac{1}{2} |\Sigma|
    \end{align}
    where $|\Sigma|$ is the determinant of the matrix $\Sigma$, and we achieve the final expression of entropy,
    \begin{align}
        H(P) = \frac{n}{2}(\log 2\pi + 1) + \frac{1}{2} |\Sigma|
    \end{align}

    Therefore, the form of the segmentation function can be written as
    \begin{align}
        &\begin{cases}
            \begin{aligned}
                &L = \frac{1}{N}|\sum_{i=1}^N \vec{v}_i| \\
                &H(P) = \log N
            \end{aligned} 
        \end{cases} & (N < \Theta),
        \\
        &\begin{cases}
            \begin{aligned}
                &L = E(\vec{v}) \\ 
                &H(P) = \frac{n}{2}(\log 2\pi + 1) + \frac{1}{2} |\Sigma|
            \end{aligned} 
        \end{cases} & (N \geq  \Theta).
    \end{align}

    Notably, this segmentation function is one of the key contributions of this paper.

    Clearly (as illustrated in Fig.~\ref{fig:t1_illustrate}), when $N < \Theta$,  entropy $H(P(I_y|I_x))$ is a monotonically increasing function of $N$. Based on the law of large numbers, as entropy increases, $N$ also increases. As a result, the mean of vector $\vec{v}$ gradually approaches the expected value, causing the distance between $\hat{I_y}$ and $D^\dagger I_x$ to decrease gradually. Notably, since the vector $\vec{v}$ is a stochastic variable, the convergence process of $\hat{I_y}$ to $D^\dagger I_x$ as $N$ increases is not strictly a Cauchy sequence \footnote{A Cauchy sequence is a sequence of numbers in which the difference between any two terms approaches zero as the sequence progresses. In other words, for any positive number $\epsilon$, there exists a term in the sequence after which all subsequent terms are within $\epsilon$ of each other.}.

    Notably, when  $N > \Theta$, it can be observed that the entropy of a Gaussian distribution is related to the determinant of the covariance matrix $\Sigma$. While an increase in entropy may lead to an increase in the covariance matrix, as $N$ approaches a sufficiently large value, 
    \begin{equation}
        L = |\sum_{i=1}^N p(\vec{v}_i) \vec{v}_i| \to E(\vec{v}) = 0,
    \end{equation}
    thus, the entropy is taken independent of $L$ and $\hat{I_y}$ will coincide with $D^\dagger I_x$. The proof is complete.
\end{proof}

\begin{theorem}\label{the:entropy-distance}{(The Entropy-Distance Correlation)} Let $\hat{I_y}$ denotes the midpoint of Euclidean distance in high-resolution distribution, $\hat{I_y}$ also can be regarded as the output of the model (Eq.~\ref{eq:model_argmin}),
    \begin{equation}
        \hat{I_y} = \mathop{\arg\min}_{I_y^\prime} \sum_{i=1}^N d(I_y^i, I_y^\prime).
    \end{equation}
    As data entropy increases, the average distance between the $\hat{I_y}$ to the high-resolution distribution will increase.
\end{theorem}

\begin{proof}
    As we proved in Eq.~\ref{eq:gd-entropy}
    \begin{equation}
        H(P(I_y|I_x))=\frac{n}{2}(\log 2\pi + 1)+\frac{1}{2}|\Sigma|,
    \end{equation}
    entropy is proportional to the covariance of the Gaussian distribution, and the uncertainty of the Gaussian distribution increases as entropy increases.

    When the number of points is large ( $N > \Theta$), the midpoint of the Euclidean distance of the Gaussian distribution can be approximated by its mean,
    \begin{equation}
        \mu = \hat{I_y} =  \mathop{\arg\min}_{I_y^\prime} \sum_{i=1}^N d(I_y^i, I_y^\prime),
    \end{equation}
    where $d(\cdot)$ is Euclidean distance. Therefore, we can calculate the mean square distance $L$ between the mean in distribution to the other points,
    \begin{equation}
        L = \frac{1}{n}\sum_{i=1}^N(I_y^i-\mu)^2.
    \end{equation}
    The mean distance between the two groups can be calculated by using the expected value formula
    \begin{align}
        E[L] &= E[(I_y^i-\mu)^2] \notag \\
        &=\text{tr}(\Sigma).\ \text{(proved in Eq.~\ref{eq:trace})}
    \end{align}

Since the expectation of $P(I_y|I_x)$ is $\mu$, we have
    \begin{align}
        E_P[||I_y-\mu||^2_2] & = E_P [(I_y - \mu)^T (I_y - \mu)] \notag \\
        & = E_P [I_y^TI_y-\mu^TI_y-I_y^T\mu+\mu^T\mu] \notag \\
        & = E_P [I_y^T I_y] - \mu^T E_P [I_y] - E_P [I_y]^T\mu + \mu^T\mu \notag \\
        & = E_P [I_y^T I_y] - \mu^T\mu - \mu^T\mu + \mu^T\mu \notag \\
        & = E_P [I_y^T I_y] - \mu^T \mu \notag \\
        &= E_P[\text{tr}(I_y^T I_y)] - \mu^T \mu \notag \\
        &= E_P[\text{tr}(I_yI_y^T)] - \mu^T \mu \notag \\
        &= \text{tr}(E_P[I_yI_y^T]) - \mu^T \mu \notag \\
        &= \text{tr}(\Sigma+ \mu\mu^T) - \mu^T \mu \label{eq:second-moment} \\
        &= \text{tr}(\Sigma) + \mu^T\mu - \mu^T \mu \notag \\
        &= \text{tr}(\Sigma) \label{eq:trace}
    \end{align}
    where $\text{tr}(\cdot)$ is the trace of the matrix and Eq.~\ref{eq:second-moment} using the second moment of Gaussian distribution. 

    As entropy increases, so does the covariance matrix. Thus, the average distance between the midpoint of Euclidean distance in the HR distribution to the HR distribution will increase. The proof is complete.
\end{proof}

These theorems provide a comprehensive explanation for how the information entropy impacts the performance of PSNR-oriented models in generating high perceptual quality images. Specifically, we demonstrate that as the entropy of the HR distribution increases, the midpoint of the Euclidean distance gradually approaches the projection point of the LR onto the HR hyperplane, eventually resulting in an overlap with the blurriest point in the HR hyperplane. Furthermore, with the increase in entropy, the mean distance between the midpoint of the Euclidean distance and the HR distribution points also increases, indicating that the generated images become further away from the true data distribution. These findings highlight the significance of entropy management in developing high-quality super-resolution models.

\subsection{Relevance with Range-Null Space Decomposition}\label{sec:rnd}
Furthermore, we can choose a specific form of $\vec{v}$ in Eq.~\ref{eq:vec_v} as $\vec{v} = (I-D^\dagger D)y_q$, where $y_q$ follows the distribution $P(I_y|I_x)$. Then we can reformulate Eq.~\ref{eq:vec_v} as 
\begin{align} 
    I_y &= D^\dagger I_x + \vec{v} \notag \\
    &= D^\dagger I_x + (I-D^\dagger D)y_q. 
\end{align} 
This equation exhibits an interesting property when we multiply both sides by $D$, yielding
\begin{equation} 
    DI_y = DD^\dagger I_x+D(I-D^\dagger D)y_q = I_x + 0, 
\end{equation} 
where $0$ denotes the zero vector. Notice that the term $D^\dagger I_x$ on the left-hand side becomes $I_x$ when we multiply it by $D$ on the left, whereas the term $(I-D^\dagger D)y_q$ on the right-hand side disappears, resulting in $0$. Therefore, the term on the left-hand side can be referred to as the range-space, which contains all possible linear combinations of $DI_y$, while the term on the right-hand side is called the null-space, which contains all possible linear combinations that lead to zero. Here, $y_q$ represents the projection of the null-space. The decomposition above is also known as the Range-Null Space Decomposition (RND).

Theorem~\ref{the:entropy-midpoint} can be reformulated in terms of the null-space. The midpoint $\hat{I_y}$ of the Euclidean distance distribution $P(I_y|I_x)$ can be expressed as 
\begin{align} 
    \hat{I_y}&=\mathop{\arg\min}_{I_y^\prime} \sum_{i=1}^N d(I_y^i, I_y^\prime) \notag \\
    &=E(P(I_y|I_x)) \notag \\
    &=\sum_{i=1}^N p(I_y^i|I_x) I_y^i \ \ \ (\text{where} \sum_{i=1}^N p(I_y^i|I_x) = 1) \notag \\
    &=\sum_{i=1}^N \left[p(I_y^i|I_x) (D^\dagger I_x + (I-D^\dagger D)y_q)\right] \notag \\
    &=D^\dagger I_x + p(I_y^i|I_x)(I-D^\dagger D)E(y_q) \notag \\ 
    &=D^\dagger I_x + p(I_y^i|I_x)(I-D^\dagger D)E(P(I_y|I_x)). 
\end{align} 
Therefore, the distance between the midpoint of the distribution $\hat{I_y}$ and the projection point on the hyperplane $D^\dagger I_x$ can be rewritten as 
\begin{align} 
    L &= |\hat{I_y} - D^\dagger I_x| \notag \\
    &= |D^\dagger I_x + p(I_y^i|I_x)(I-D^\dagger D)E(P(I_y|I_x)) - D^\dagger I_x| \notag \\
    &= |p(I_y^i|I_x)(I-D^\dagger D)E(P(I_y|I_x))|. 
\end{align} 
Since $(I-D^\dagger D)$ and $E(P(I_y|I_x))$ are constant, let $c=(I-D^\dagger D)E(P(I_y|I_x))$ denote a constant, the distance thus becomes 
\begin{equation} 
    L = |c|p(I_y^i|I_x). 
\end{equation}
As demonstrated in Sec~\ref{sec:entropy}, the information entropy is positively influenced by two main factors: the variation of the distribution and the number of data. Using RND, we can directly examine the relationship between distance $L$ and the entropy of the distribution $H(P(I_y|I_x))$. This provides a more abstract interpretation that does not rely on the assumption that the distribution $P(I_y|I_x)$ conforms to a Gaussian distribution.

\begin{figure}[!tb]
\centering
\includegraphics[width=\linewidth]{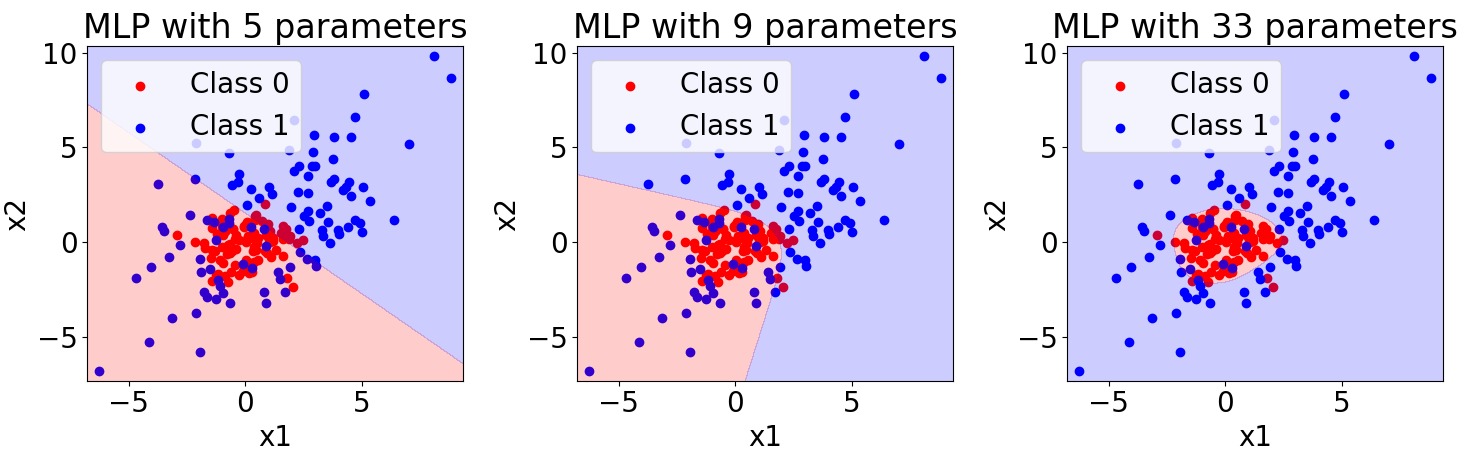}
\caption{A toy example of how model size influences the discriminative power. Increasing the number of parameters of the Multi-layer Perceptron model (MLP) improves its discriminative power by making the decision boundary more flexible and adaptive to the data distribution.}
\label{fig:toy}
\end{figure}

\begin{figure*}[!t]
  \centering
  \includegraphics[width=1.0\linewidth]{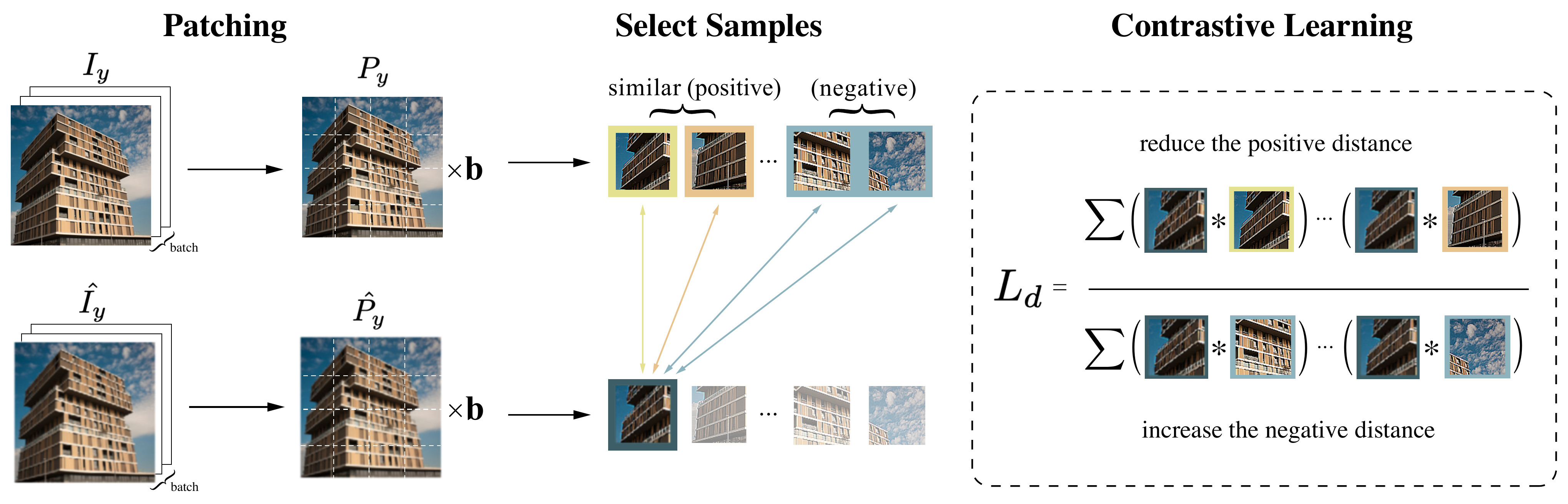}
  \caption{A schematic overview of the DECLoss method. The generated images $\hat{I_y}$ and HR images $I_y$ are first divided into a sequence of flattened patches $\hat{P_y}$ and $P_y$, respectively. Then, patches are classified into positive and negative samples based on their similarities. Finally, contrastive loss $L_d$ is performed on the patches to polarize the details thus reducing the impact of the COO problem ($*$ is an operator represents Eqs.~\ref{eq:cl_sr2hr} to \ref{Eq_Q_neg}).}
  \label{fig:decloss_overview}
\end{figure*}

\subsection{Discussion}
Numerous recent studies have endeavored to tackle the issue of over-smoothing and proposed a diverse range of techniques to produce high-quality images with enhanced perceptual features. These approaches can be broadly classified as attempting to implicitly optimize the COO problem from two different viewpoints.

Approaches of the first perspective aim to enhance the model's ability to learn the conditional distribution by optimizing network blocks or increasing the model's number of parameters, such as SwinIR \cite{swin} and RCAN \cite{rcan}. By improving model capacity, the discriminative ability towards similar low-resolution inputs also increases, leading to a reduction in the potential range of values for the high-resolution conditional distribution given a specific low-resolution input. As a result, the entropy $H(I_y|I_x)$ is decreased, and the output can be closer to the HR distribution (Theorem~\ref{the:entropy-distance}).
However, such methods typically require numerous experiments and consume significant computational resources, making them computationally inefficient.

Note that, ideally, the model does not alter the entropy of the data, but in practice, the entropy of the data is related to the discriminative power of a model. This is because it is unlikely to encounter situations where different high-resolution images have exactly the same low-resolution counterparts. Generally, the more capable the model is of distinguishing similar low-resolution images, the fewer possible high-resolution images there will be, and consequently, the lower the entropy. To illustrate this point more clearly, we present a toy example as shown in Fig.~\ref{fig:toy}. We train 3 Multi-layer Perceptrons (MLP) with different units, each MLP has an input layer, a hidden layer, and an output layer. We provide more details of this toy example in Appendix D.

In the second perspective of SR approaches, researchers have explored an alternative approach by optimizing the distance between distributions instead of relying solely on the Euclidean distance. These approaches involve transforming SR into a conditional generation problem and utilizing various generative frameworks, such as GAN-based loss \cite{esrgan, ranksrgan}, flow-based models \cite{srflow}, and diffusion-based models \cite{sr3}, to achieve higher perceptual quality. In addition, the Perceptual Loss \cite{perceptual_loss} approach employs a pre-trained neural network to constrain the feature distance between generated images and real images. The feature distance represents the perceptual quality distance, and the neural network loss is a relaxed version of the Euclidean distance loss. Despite these advantages, the results generated by these models often exhibit a high degree of randomness, leading to poor consistency with the input.

To address this issue, we have proposed an explicit solution approach that involves reducing the variance of the potential HR distribution. Specifically, we propose the \emph{Detail Enhanced Contrastive Loss (DECLoss)}, which utilizes the clustering property of contrastive loss to alleviate the problem of high entropy in SR generation. By reducing the variance of the HR distribution, DECLoss aims to produce higher-quality SR images with improved consistency.


\section{Method}\label{sec:method}
\subsection{Motivation}
In Fig.~\ref{fig:coo_problem} (d), we present an approach for estimating the distribution of the high-resolution (HR) image conditioned on a low-resolution (LR) image $P(I_y|I_x)$, where we assume that this distribution follows a Gaussian distribution. In essence, we aim to reduce the variance of this distribution by increasing the probabilities of a few HR pixels close to the mean and decreasing the probabilities of the remaining HR pixels. 
Note that, we are not referring to the property of the distribution of generated high-resolution images given a low-resolution image, but that of the potential high-resolution images given a low-resolution image. 
To this end, we propose utilizing a contrastive learning method \cite{simclr}. However, applying contrastive learning at the pixel-level is a challenging task due to its exponential computational requirements and difficulty in setting positive and negative samples. Therefore, we propose to relax the constraints to the patch-level.

It is important to note that in our proposed approach, we assume that small patches possess the same properties as individual pixels and remain consistent with the characteristics of the COO problem. Specifically, we interpret the center of the circle in Fig.~\ref{fig:coo_problem} (d) as a set of LR patches with differences smaller than the model's discriminant lower bound, whereas the arc represents the potential HR patches. This patch-level approach provides a suitable framework for achieving the desired variance reduction.


\subsection{Detail Enhanced Contrastive Loss}

\subsubsection{Patching}
We begin our approach by relaxing the constraints from the pixel-level to the patch-level. This involves reshaping the high-resolution (HR) image $I_y$ and its corresponding predicted image $\hat{I_y}$, both of dimensions $\mathbb{R}^{B\times C\times H\times W}$, into a sequence of flattened 2D patches $P_y$ and $\hat{P_y}$, respectively. The shape of each patch is given by $(P, P)$, where $P$ is the patch resolution, and $N = HW/P^2$ represents the total number of patches in each image. This approach facilitates the application of contrastive learning in selecting positive and negative samples, which is essential to our method.

\subsubsection{Selection of Positive and Negative Samples}
Regarding the selection of positive and negative samples, we adopt the PSNR similarity metric to separate them. Specifically, we define positive samples as those with a PSNR similarity measure greater than or equal to a specified threshold $\eta$, while negative samples correspond to those with a PSNR similarity measure less than $\eta$. The PSNR similarity measure $M_{i,j}$ for a pair of patches $P_y^i$ and $P_y^j$ is defined as
\begin{equation}\label{eq:mask_thres}
    M_{i,j} = -20\times log_{10}(\frac{||P_y^i-P_y^j||_2}{\text{MAX}}), M \in \mathbb{R}^{B_p\times B_p},
\end{equation}
where $||.||_2$ denotes the $L_2$-norm, $B_p = B \times N$, and $\text{MAX}$ represents the upper bound of the color space. Furthermore, when the PSNR similarities of corresponding HR patches are greater than the threshold $\eta$, we treat the corresponding LR patches as one patch. This approach is preferred because PSNR is an easy-to-set threshold metric that effectively discriminates between positive and negative samples.

\subsubsection{Contrastive Learning}
Taking inspiration from \cite{simclr}, we employ the cosine similarity as the constraint term. To ensure the stability of the constraint, we measure the cosine similarities not only from the predicted high-resolution patches $\hat{P_y}$ to the ground truth high-resolution patches $P_y$ (referred to as $S_{sr \rightarrow hr}$), but also from $\hat{P_y}$ to $\hat{P_y}$ (referred to as $S_{sr \rightarrow sr}$). Mathematically, the cosine similarities are computed as follows,
\begin{equation}
    S^{i,j}_{sr \rightarrow hr}=\frac{P_y^i\hat{P}_y^j}{|P_y^i||\hat{P}_y^j|}, S_{sr \rightarrow hr} \in \mathbb{R}^{B_p \times B_p},
\label{eq:cl_sr2hr}
\end{equation}

\begin{equation}\label{eq:cl_sr2sr}
    S^{i,j}_{sr \rightarrow sr}=\frac{\hat{P}_y^i\hat{P}_y^j}{|\hat{P}_y^i||\hat{P}_y^j|}, S_{sr \rightarrow sr} \in \mathbb{R}^{B_p \times B_p}.
\end{equation}
where $P_y^i$ and $\hat{P}_y^j$ represent the $i$-th and $j$-th patches of $P_y$ and $\hat{P_y}$, respectively, and we measure the similarities of samples across the current batch $B_p = B \times N$.

A desirable property of the contrastive samples is that they should originate from similar/dissimilar patches with comparable textures across the entire dataset. However, this is computationally prohibitive, as it would entail loading the entire dataset and calculating the similarity/dissimilarity of patches for each training iteration. Hence, we adopt the batch-wise strategy, which is a prevalent practice in many machine-learning methods. When the batch size is sufficiently large, it can be regarded as a reasonable compromise between the computational cost and the approximation of the entire dataset.

Next, we compute the scores for the positive and negative samples, represented as $Q_{pos}$ and $Q_{neg}$, respectively. Specifically, for the positive samples, we sum over all $j$ such that the PSNR similarity $M_{ij}$ of the $i$-th and $j$-th patches of $P_y$ is greater than or equal to the threshold $\eta$. The score is computed as a function of the cosine similarities as 
\begin{equation}
   Q_{pos}^i = \sum_{j=1}^{B_p} \mathbbm{1}_{[M_{ij} \ge \eta]} \left[\exp({S_{sr \to sr}^{ij}})+2\exp({S_{sr \to hr}^{ij}})\right]/t_{pos},
\end{equation}
where $t_{pos}$ is a temperature parameter. Similarly, for the negative samples, we sum over all $j$ such that $M_{ij}$ is less than $\eta$, and compute the score
\begin{equation}
   Q_{neg}^i = \sum_{j=1}^{B_p} \mathbbm{1}_{[M_{ij}<\eta]} \left[\exp({S_{sr \to sr}^{ij}})+2\exp({S_{sr \to hr}^{ij}})\right]/t_{neg}.
\label{Eq_Q_neg}
\end{equation}

Finally, we define the Detail Enhanced Contrastive Loss (DECLoss), denoted as $L_d$,
\begin{equation}\label{eq:decloss}
    L_d = -\frac{1}{B_p}\sum_{i=1}^{B_p}\log (\frac{Q_{pos}^i}{Q_{neg}^i}).
\end{equation}

\subsection{Applying DECLoss}
We propose an algorithm shown in Alg.~\ref{alg_training} for training a super-resolution model $F_\theta$ with DECLoss. In Alg.~\ref{alg_training}, we combine DECLoss with $L_1$ loss and perceptual loss to enhance the image reconstruction quality as used in ESRGAN \cite{esrgan}.
The $L_1$ loss is defined as the 1-norm distance between the generated image and its ground truth, which ensures that the reconstructed image has high fidelity to the original. The $L_1$ loss is computed as
\begin{equation}\label{eq:l1_loss}
    L_1 = ||\hat{y}- y||_1.
\end{equation}
where $\hat{y}$ and $y$ are the generated and ground truth images, respectively.
In addition to the $L_1$ loss, the perceptual loss is used to measure the distance between the feature representations of the generated and ground truth images. A pre-trained VGG-19 network \cite{vgg} is used to generate a feature map $\varphi$, which is used to compute the perceptual loss. Specifically, the perceptual loss is defined as
\begin{equation}\label{eq:percep_loss}
    L_p = ||\varphi(\hat{y})-\varphi(y)||_2.
\end{equation}

\begin{table*}[t]
\caption{Ablation study results. We compare different configurations of loss functions. We calculate restoration metrics PSNR and perceptual metric LPIPS \mbox{
\cite{lpips}}\hspace{0pt}
. Config. denotes different loss configurations. $L_p$ denotes the perceptual loss \mbox{
\cite{perceptual_loss} }\hspace{0pt}
and RaGAN is the GAN metric of ESRGAN \mbox{
\cite{esrgan}}\hspace{0pt}
. All metrics are calculated on the DIV2K \mbox{
\cite{div2k} }\hspace{0pt}
validation set.}
\label{tab:ablation}
\centering
\scalebox{1}{
\begin{tabular}{cccccccc} 
\hline
Config. & Model & L1 & $L_p$ (Eq.~\ref{eq:percep_loss}) & $L_d$ (Eq.~\ref{eq:decloss}) & RaGAN & $\uparrow$ PSNR & $\downarrow$ LPIPS  \\ 
\hline
1     & RRDB  & \checkmark  &    &     &        & 29.47    & 0.130     \\ 
2 & RRDB  & \checkmark  &    &  \checkmark   &        & 29.17    & 0.121     \\ 
3 & RRDB  & \checkmark  &  \checkmark  &     &        & 29.10    & 0.117     \\ 
4   & RRDB  & \checkmark  & \checkmark  & \checkmark     &        & 28.87    & \textbf{0.108}   \\
\hline
5 & RRDB  & \checkmark  & \checkmark   &     & \checkmark    & 26.63    & 0.092 \\
6  & RRDB  & \checkmark  & \checkmark   & \checkmark   &   \checkmark   & 28.19    & \textbf{0.080} \\
\hline
\end{tabular}}
\end{table*}

\begin{figure*}[!t]
\centering
\includegraphics[width=\linewidth]{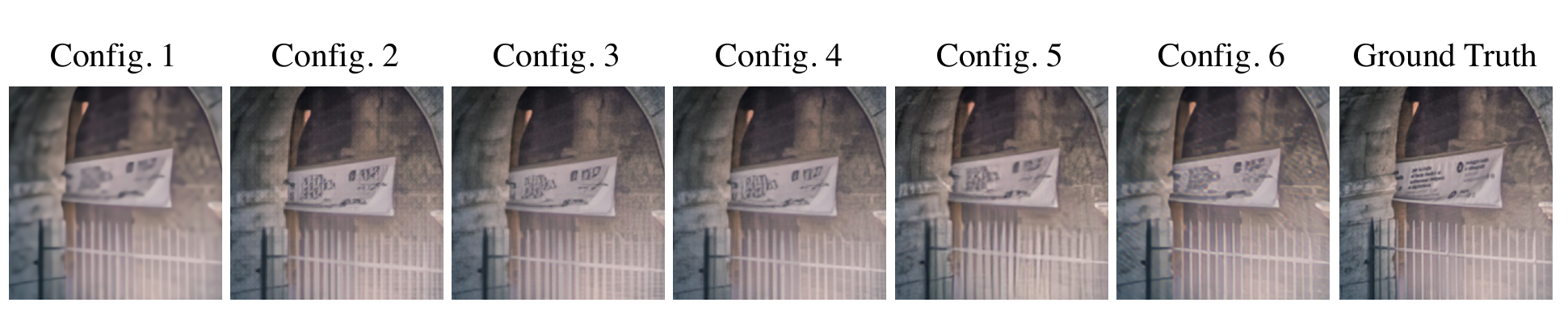}
\caption{Comparison of different configurations of losses.}
\label{fig:ablation}
\end{figure*}

The final loss function $L$ is a weighted sum of the $L_1$, perceptual, and the Detail Enhanced Contrastive loss (DECLoss), which is defined as 
\begin{equation}\label{eq:final_loss}
    L = \omega_1 \times L_1 + \omega_2 \times L_p + \omega_3 \times L_d.
\end{equation}
The weights $\omega_1$, $\omega_2$, and $\omega_3$ are used to balance the contributions of each loss term and are determined empirically. The DECLoss loss $L_d$ is defined in Eq.~\eqref{eq:decloss}, where the distance between $Q_{pos}^i$ and $Q_{neg}^i$ is constrained such that the distance of $Q_{pos}^i$ is reduced if the distance of $Q_{neg}^i$ is increased. This encourages the clustering of positive samples, which aids in the enhancement of image features.

\begin{algorithm}[ht]
    \caption{Applying DECLoss to train a super-resolution model}
    \renewcommand{\algorithmicrequire}{\textbf{Input:}}
    \renewcommand{\algorithmicensure}{\textbf{Output:}}
    \label{alg_training}
    \begin{algorithmic}[1]
    \REQUIRE LR image dataset $\mathcal{X}$ and corresponding HR image dataset $\mathcal{Y}$, number of training iteration $T$.
    \ENSURE Super-resolution model $F_\theta$.  

    \FOR{ $t = 1, 2, \cdots, T $}
        \STATE{Sample a batch of LR images $I_x$ and corresponding HR images $I_y$ from $\mathcal{X}$ and $\mathcal{Y}$, respectively. }
        \STATE{Obtain SR images: $\hat{I}_y = F_\theta(I_x)$.}
        \STATE{Reshape $I_y$ and $\hat{I}_y$ into patches $P_y$ and $\hat{P}_y$, respectively.}
        \STATE{Obtain positive samples $Q_{pos}$ and negative samples $Q_{neg}$ using $P_y$ and $\hat{P}_y$ as Eqs.~\ref{eq:cl_sr2hr} to \ref{Eq_Q_neg}.}
        \STATE{Calculate the DECLoss using positive samples $Q_{pos}$ and negative samples $Q_{neg}$ as Eq.~\ref{eq:decloss}. }
        \STATE{Calculate the $L_1$ loss and perceptual loss using Eqs.~\ref{eq:l1_loss} to \ref{eq:percep_loss}. }
        \STATE{Calculate the final loss $L$ using DECLoss, $L_1$ loss and perceptual loss as Eq.~\ref{eq:final_loss}.}
        \STATE{Update the parameter of $F_\theta$ by taking a stochastic gradient descent step on $L$.}
    \ENDFOR
    \end{algorithmic}
\end{algorithm}

Notably, the COO problem is a fundamental challenge in image super-resolution, which arises from the ill-posed nature, and is a consequence of employing MSE Loss \cite{srresnet, pd}, which inherently struggles to capture the full complexity of real-world image degradation and restoration processes \cite{beby_gan}. As a result, achieving a complete resolution of the COO problem becomes an intricate endeavor \cite{pd}. Our approach offers a noteworthy step forward by providing a partial resolution to the COO problem. Through extensive experimentation and careful consideration, we've endeavored to navigate the delicate balance between preserving the truthfulness of the restored images and enhancing their perceptual consistency. Our method strives to achieve the best attainable trade-off within the constraints of the current framework. It's important to emphasize that while our method represents a significant advancement in the realm of super-resolution, a complete resolution to the COO problem is indeed a formidable challenge.

\section{Experimentation}\label{sec:experiment}
\subsection{Settings} 
The models are trained on two large-scale datasets, DIV2K \cite{div2k} and Flickr2K \cite{edsr}, using bicubic interpolation to down-sample the high-resolution (HR) images and obtain corresponding low-resolution (LR) training data. The trained models are evaluated using standard super-resolution (SR) benchmarks, including DIV2K(test) \cite{div2k}, BSD100 \cite{bsd100}, and Urban100 \cite{urban100}. Specifically, we focus on evaluating the performance of the RRDB \cite{esrgan} and SwinIR \cite{swin} models through a series of experiments. Consistent with ESRGAN \cite{esrgan}, we set the scaling factor between LR and HR images to $\times 4$. In addition, patches of size $32 \times 32$ and $128 \times 128$ are extracted from LR and HR images, respectively, to enable fair comparisons across different models. To ensure stable and efficient training, we employ a batch size of 32 and train each model for 1,000 batches per epoch.

The training process is a crucial component of deep learning models, as it is responsible for fine-tuning the weights and biases of the network to optimize performance. In this study, we divided the training process into two stages to ensure optimal results. In the first stage, all models were pre-trained with the $L_1$ loss function (Eq.~\ref{eq:l1_loss}) for 200 epochs, with 5 warm-up epochs. We employed a cosine learning rate with a decay rate of $\gamma=0.5$ and an initial learning rate of $1\times 10^{-4}$. This pre-training step with $L_1$ loss facilitates the model to converge to an appropriate range.
In the second stage, we trained the generator with the loss function defined in Eq.~\eqref{eq:final_loss}, where $\omega_1=1\times 10^{-2}, \omega_2 = 1$, and $\omega_3=1\times 10^{-2}$, due to the small real value of perceptual loss $L_p$ and the generally large value of DECLoss $L_d$. We set the temperature $(t_{pos}, t_{neg})$ to $(0.5, 1.5)$, and we used a patch size of $8\times 8$, given the trade-off between the number of patches and computational resources. The learning rate was set to $2\times 10^{-5}$, and the cosine learning rate was decayed according to a schedule. We employed the Adam algorithm \cite{adam} for optimization, with $\beta_1=0.9$ and $\beta_2 = 0.999$. The models were implemented using PyTorch and trained on 4$\times$ A100 GPUs to ensure computational efficiency.

\subsection{Ablation studies}
\label{subsec:ablation}

\begin{table}[t]
\centering
\caption{Influence of different thresholds of distinguishing the positive and negative samples.}
\label{tab:threshold}
\begin{tabular}{ccc}
\hline
Threshold     & $\uparrow$ PSNR   & $\downarrow$ LPIPS    \\
\hline
30  & 27.854 & 0.1634  \\
32  & 27.859 & 0.1630  \\
34  & 27.860 & \textbf{0.1626} \\
36 & 27.843 & 0.1701  \\
38 & 28.046 & 0.1754  \\
40 & \textbf{28.449} & 0.1830  \\
\hline
\end{tabular}
\end{table}

\begin{table}[t]
\centering
\caption{Comparison of a different combination of batch size with patch size.}
\label{tab:patchsize}
\begin{tabular}{ccc}
\hline
batch/patch     & $\uparrow$ PSNR   & $\downarrow$ LPIPS \\
\hline
32/4  & 27.648 & 0.1645 \\
28/6  & 27.895 & 0.1648 \\
24/8  & 27.860 & \textbf{0.1602} \\
16/12 & 28.021 & 0.1713  \\
12/16 & \textbf{28.230} & 0.1809  \\
\hline
\end{tabular}
\end{table}

\subsubsection{Comparisons of different loss configurations.} 
To demonstrate the effectiveness of our proposed DECLoss, we conducted an ablation study with different loss configurations. Table~\ref{tab:ablation} and Fig.\ref{fig:ablation} summarize the results of the study. Config. 1 represents the baseline RRDB \cite{esrgan} trained with $L_1$ loss, while Config. 2 shows the results of RRDB trained with both $L_1$ loss and our DECLoss. Config. 3 is the results trained by $L_1$ loss and Perceptual Loss \cite{perceptual_loss}. Config. 4 corresponds to RRDB trained with our proposed DECLoss (Eq.\ref{eq:decloss}). Config. 5 represents the official results of ESRGAN, where RaGAN is combined with RRDB. Finally, Config. 6 demonstrates the results of our proposed DECLoss combined with RaGAN.

Our experimental results indicate that Config. 6 achieves higher subjective and objective quality compared to the ESRGAN (Config. 5). Specifically, the incorporation of DECLoss (Config. 4) and the combination of DECLoss with RaGAN (Config. 6) show notable improvements in both subjective and objective evaluations compared to Config. 1, Config. 3 and Config.  5. Furthermore, adding DECLoss to L1 loss alone (Config. 2) also leads to a significant improvement in the LPIPS score. Based on these observations, we conclude that our method is a simple yet effective way to enhance the perceptual quality of the super-resolved images. These findings demonstrate the efficacy and robustness of our proposed DECLoss.

\begin{figure}[!t]
    \centering
    \includegraphics[width=0.9\linewidth]{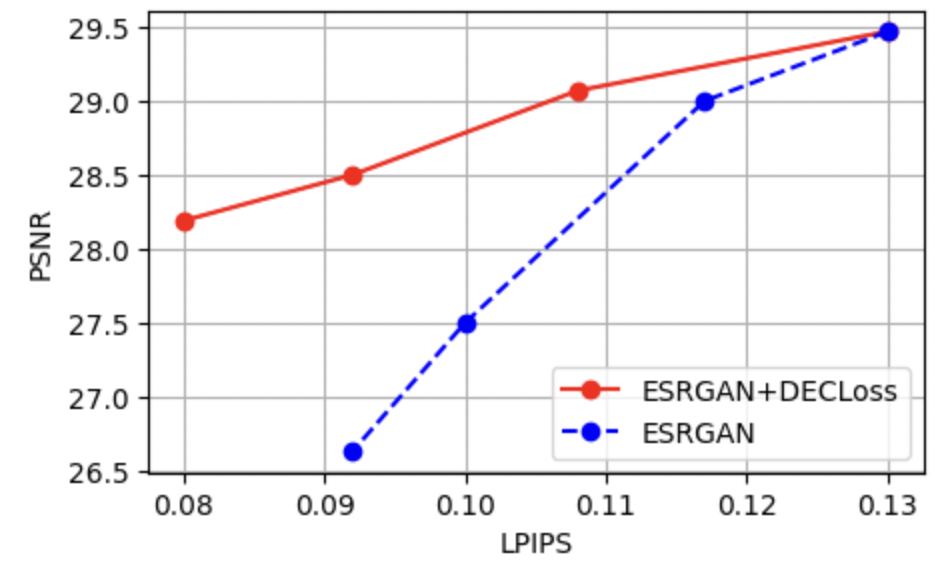}
    \caption{The Pareto front: comparisons of ESRGAN with or without DECLoss.}
    \label{fig:pareto}
\end{figure}

\begin{figure}[t]
\centering
\includegraphics[width=0.9\linewidth]{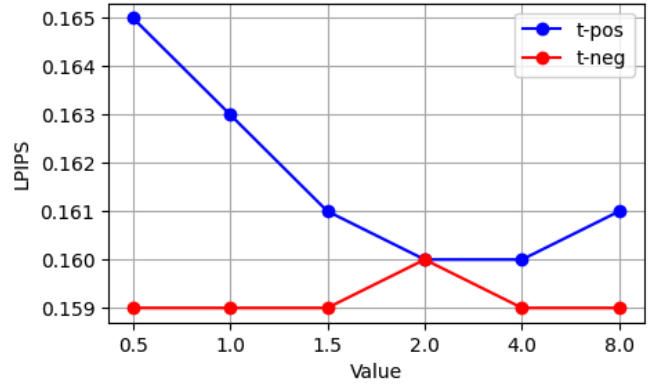}
\caption{Comparisons of different temperatures. The blue curve represents the LPIPS variation of $t_{pos}$, and the red curve represents $t_{neg}$.}
\label{fig:temperature}
\end{figure}

In order to investigate the perception-distortion tradeoff of our proposed method, we conduct ablation studies with different settings of ESRGAN with or without DECLoss, and plot the tradeoff curve on the DIV2K \cite{div2k}
dataset. Specifically, we use two extreme cases as the baseline: using only L1 loss as the reconstruction loss and using all losses as the full objective function, respectively, and we interpolate between them by varying the weights of the losses to obtain different tradeoff points. We use the same metrics as before to evaluate the performance of the methods: PSNR and LPIPS. The results are shown in Figure \ref{fig:pareto}. As can be seen, our method with DECLoss achieves a new Pareto front of the tradeoff curve, meaning that it outperforms ESRGAN without DECLoss on both metrics for any given tradeoff point. We acknowledge that the trade-off between perception and distortion remains a complex matter, and we firmly believe that each model possesses its own ideal perception-distortion curve. Our method endeavors to approach this ideal by optimizing the trade-off to create images with enhanced perceptual quality while minimizing distortion.

\begin{table*}[!t]
\centering
\caption{Comparisons with state-of-the-art SR methods. Specifically, DECLoss is combined with RaGAN and trained on RRDB \cite{esrgan}. }
\label{tab:sota}
\scalebox{1}{
\begin{tabular}{ccccccccccc} 
\hline
Benchmark                 & Metric & RankSRGAN \cite{ranksrgan} & ESRGAN \cite{esrgan} & SwinIR-Real \cite{swin}  & SPSR \cite{spsr:ma2021structure} & LDL \cite{ldl:liang2022details} & FeMaSR \cite{femasr:chen2022real} & \textbf{DECLoss} \\ 
\hline
\multirow{2}{*}{DIV2K}    & $\uparrow$ PSNR  &    26.55      & 26.64 & 27.80  & 26.70 & 27.32 & 22.96 & \textbf{28.19} \\
                          & $\downarrow$ LPIPS  &    0.099     & 0.092   & 0.105  & 0.091 & 0.082 & 0.158 & \textbf{0.080} \\
\hline
\multirow{2}{*}{Urban100} & $\uparrow$ PSNR    &  22.98       &   22.78  &  23.08  & 23.24 & 23.90 & 20.21 & \textbf{24.51} \\
                          & $\downarrow$ LPIPS   &  0.122      &   0.108   &  0.103 & 0.104 & 0.094 & 0.165 & \textbf{0.093} \\
\hline
\multirow{2}{*}{BSD100}   & $\uparrow$ PSNR  &   24.13  &    23.97 & 24.57   & 24.12 & 24.39 & 21.77 & \textbf{24.61} \\
                          & $\downarrow$ LPIPS &  0.114   & 0.108 &  0.110 & 0.109 & \textbf{0.102} & 0.157 & 0.105 \\
\hline
\end{tabular}
}
\end{table*}

\begin{figure*}[t]
\centering
\includegraphics[width=\linewidth]{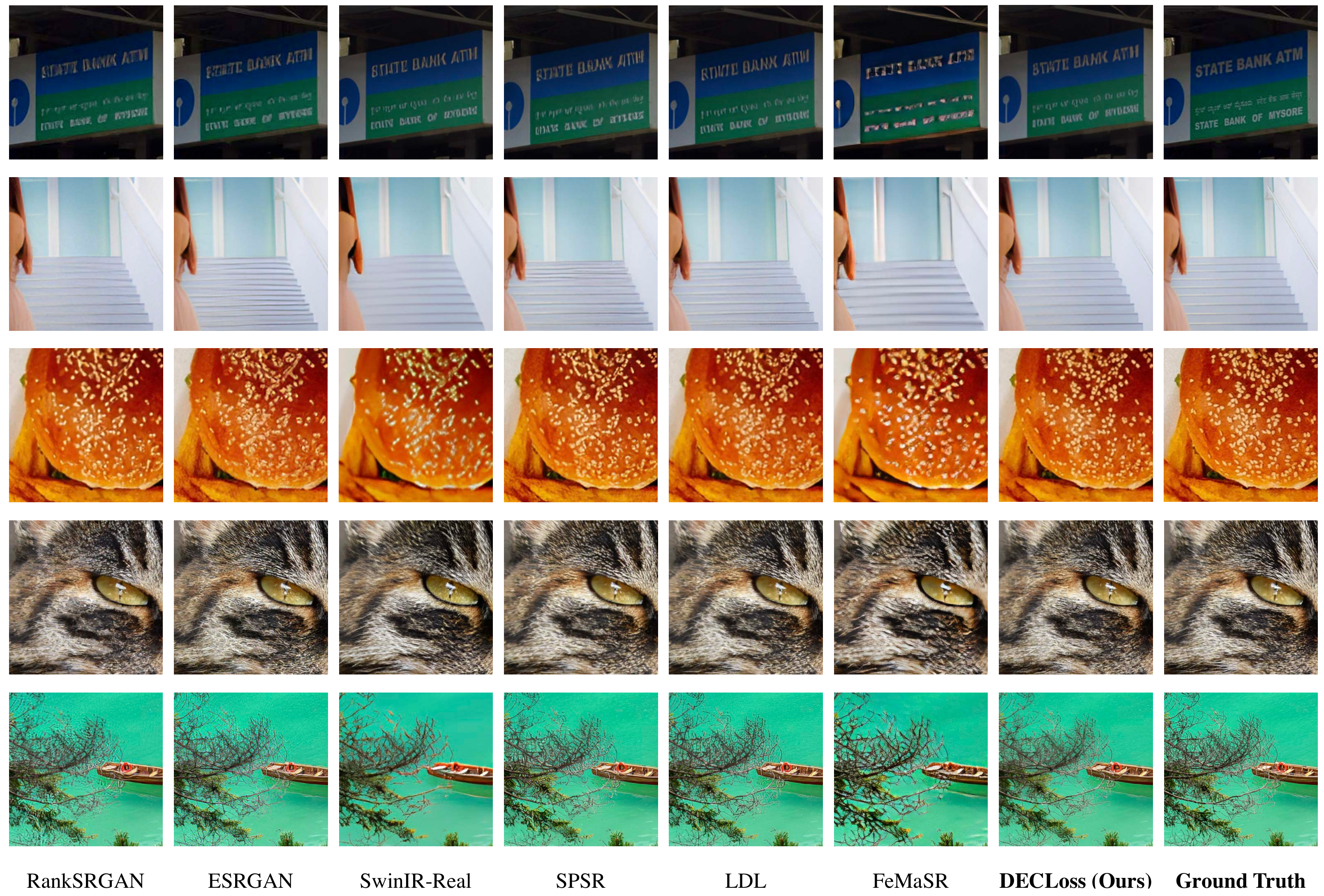}
\caption{Comparisons with state-of-the-art perceptual-driven SR methods \textbf{(zoom in for the best view)}.}
\label{fig:sota}
\end{figure*}

\subsubsection{Influence of clustering thresholds.} 
The present study aims to investigate the impact of the clustering threshold ($\eta$) on the performance of our proposed method. In particular, we focused on comparing the performance of six different values of $\eta$ ranging from 30 to 40 to determine the optimal threshold for differentiating between positive and negative samples as defined in Eq. \ref{eq:mask_thres}. Our research findings, as presented in Table \ref{tab:threshold}, demonstrate that the best performance is achieved with a threshold of 34. In other words, patches with similarities greater than this threshold value are considered positive while the remaining ones are negative. However, our results also show that the performance remains stable within a certain range of values (i.e., 30-36), indicating the robustness of our DECLoss. Nevertheless, as the threshold value increases beyond this range, the DECLoss will gradually converge to the Euclidean distance loss, resulting in a reduction in perceptual quality.

\subsubsection{Influence of contrastive temperatures.} 
Regarding the contrastive temperatures, we conducted two sets of experiments using fixed $t_{neg}=1.5$ and varying $t_{pos}=[0.5, 1.0, 1.5, 2.0, 4.0, 8.0]$, we also fixed $t_{pos}=0.5$ and varying $t_{neg}=[0.5, 1.0, 1.5, 2.0, 4.0, 8.0]$, to evaluate the impact of different temperature values on the learning efficiency. As depicted in Fig.~\ref{fig:temperature}, our results indicate that the choice of $t_{pos}$ and $t_{neg}$ within a certain range does not have a significant effect on the performance. However, the optimal value for $t_{pos}$ was found to be $0.5$, suggesting that larger temperatures could increase the learning efficiency in the context of SR's contrastive learning.

\subsubsection{Influence of different batch and patch sizes.} 
In terms of model performance and computation consumption, the batch size and patch size are two key hyper-parameters that need to be carefully selected. While increasing these parameters may enhance model performance, it also comes at the cost of increased computational resources. Thus, finding the optimal trade-off between these parameters is crucial. In this study, we explored different combinations of batch/patch sizes, namely [32/4, 28/6, 24/8, 16/12, 12/16], each occupying a similar amount of GPU memory. The results presented in Table~\ref{tab:patchsize} indicate that the optimal combination for PSNR is 28/6, while for LPIPS it is 24/8. It is important to note that increasing the batch and patch sizes beyond a certain range may not have a significant effect on the experimental results. Our findings demonstrate the stability of our method under different conditions, which can guide the selection of hyper-parameters for similar tasks in the future.

\subsection{Comparisons with State-of-the-art Methods}
We compare our proposed method, DECLoss, with several state-of-the-art perceptual-driven SR methods on three benchmark datasets: DIV2K \cite{div2k}, Urban100 \cite{urban100}, and BSD100 \cite{bsd100} in the case of $4\times$ bicubic downsampling. We use two metrics to evaluate the performance of the methods: PSNR and LPIPS. PSNR measures the peak signal-to-noise ratio between the SR image and the ground truth image, while LPIPS measures the perceptual similarity between them. Higher PSNR and lower LPIPS indicate better quality of the SR image. The results are shown in Table~\ref{tab:sota}. As can be seen, DECLoss achieves the best performance on both metrics on the DIV2K and Urban100 datasets and the best PSNR on the BSD100 dataset. This demonstrates that DECLoss can effectively enhance the details and textures of the SR images while preserving their naturalness and realism (illustrated in Fig.~\ref{fig:sota}). In particular, DECLoss achieves 28.19 PSNR with 0.080 LPIPS in the DIV2K validation set, which has a 1.55 dB improvement in PSNR and 13\% reduction in LPIPS than ESRGAN \cite{esrgan}.
The only exception is that LDL \cite{ldl:liang2022details} has a slightly lower LPIPS than DECLoss on the BSD100 dataset, but it also has a lower PSNR than DECLoss, indicating a trade-off between fidelity and perceptual quality. However, the resolution of the BSD100 dataset is the smallest one among these three datasets, and our proposed method is more focused on large resolution. This means that DECLoss can handle more challenging SR scenarios and produce more realistic and detailed SR images.

\begin{table*}[!t]
\caption{Comparisons with state-of-the-art real-world SR methods.}
\label{tab:realsr}
\centering
\begin{tabular}{lcccccc}
\hline
                          & \multicolumn{2}{c}{RealSR Nikon}       & \multicolumn{2}{c}{RealSR Canon}       & \multicolumn{2}{c}{DRealSR}  \\ \cline{2-7}
\multirow{-2}{*}{Methods} & $\uparrow$ PSNR & $\downarrow$ LPIPS & $\uparrow$ PSNR     & $\downarrow$ LPIPS  & $\uparrow$ PSNR & $\downarrow$ LPIPS \\
\hline
Real-ESRGAN \cite{wang2021realesrgan} & 24.15 & 0.1671 & 24.46 & 0.1630 & 25.82 & 0.1473 \\
Real-ESRGAN+LDL \cite{ldl:liang2022details} & 23.84 & 0.1679 & 24.16 & 0.1477 & 25.67 & 0.1493 \\
Real-ESRGAN+\textbf{DECLoss}       & \textbf{24.19}                     & \textbf{0.1645} & \textbf{24.51}  & \textbf{0.1432} & \textbf{26.09}                         & \textbf{0.1451} \\
\hline
BSRGAN \cite{bsrgan:zhang2021designing} & 24.58 & 0.1597 & 25.18 & 0.1403 & 26.19 & \textbf{0.1493} \\
BSRGAN+\textbf{DECLoss}            & \textbf{24.73}                         & \textbf{0.1589} & \textbf{25.32} & \textbf{0.1363} & \textbf{26.38}                         & 0.1495 \\
\hline
\end{tabular}
\end{table*}

\begin{figure*}[t]
\centering
\includegraphics[width=0.9\linewidth]{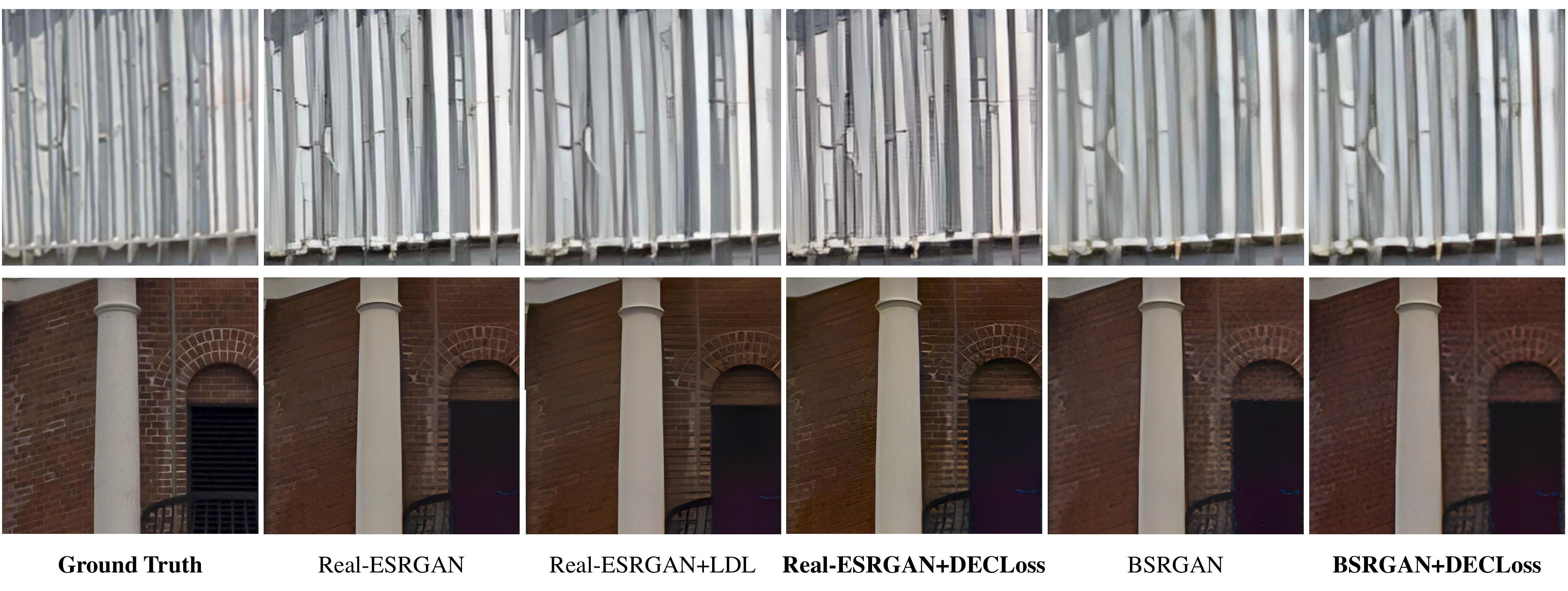}
\caption{Comparisons with state-of-the-art real-world SR methods \textbf{(zoom in for the best view)}.} 
\label{fig:realsr}
\end{figure*}

\begin{figure}[htb]
\centering
\includegraphics[width=\linewidth]{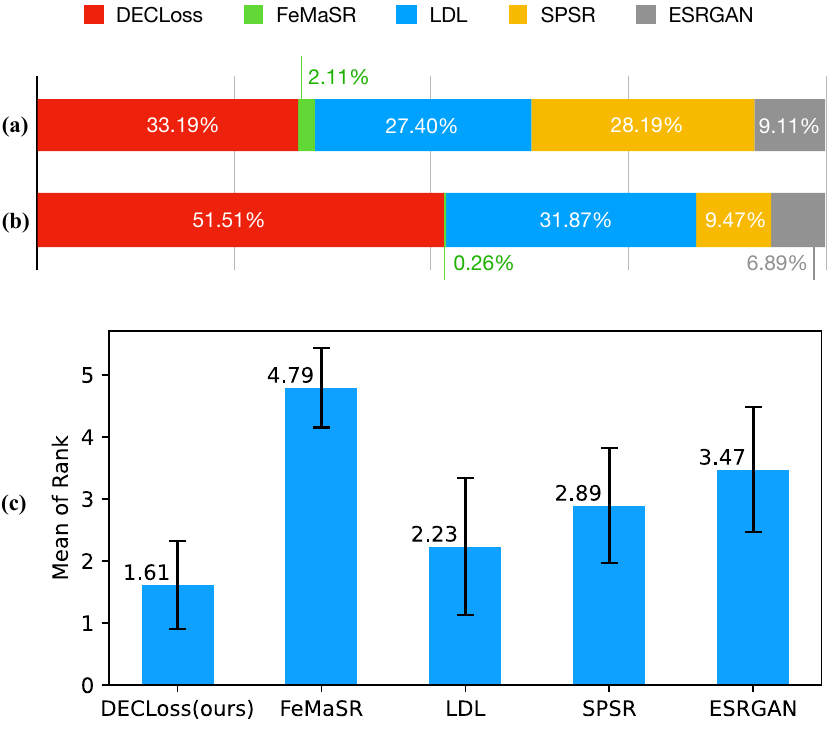}
\caption{The results of user study, comparing DECLoss (ours) with FeMaSR \cite{femasr:chen2022real}, LDL \cite{ldl:liang2022details}, SPSR \cite{spsr:ma2021structure} and ESRGAN \cite{esrgan}. The chart shows that our method achieves significantly higher user preference than the existing methods in terms of (a) percentage of top-3 images, (b) percentage of top-1 images by majority of votes, and (c) average rank of each method (the lower the better), where the black lines show the standard error of the mean.}
\label{fig:user_study}
\end{figure}

We further compare our proposed method, DECLoss, with state-of-the-art real SR methods on three real-world SR datasets: RealSR Nikon \cite{realsr-cai2019toward}, RealSR Canon \cite{realsr-cai2019toward}, and DRealSR \cite{drealsr-wei2020component}. These datasets contain real low-resolution images captured by different cameras and corresponding high-resolution images obtained by different methods. We use the same metrics as before to evaluate the performance of the methods: PSNR and LPIPS. The results are shown in Table \ref{tab:realsr}. As can be seen, DECLoss improves the performance of both Real-ESRGAN \cite{wang2021realesrgan} and BSRGAN \cite{bsrgan:zhang2021designing} on all three datasets on both metrics. This shows that DECLoss can handle real-world SR problems and generate more realistic and detailed SR images. We also show some visual examples of the SR images generated by DECLoss and other methods on real-world SR datasets in Fig.~\ref{fig:realsr} to illustrate the difference in quality, demonstrating our effectiveness and robustness.

\subsection{User Study}

We conducted a user study to evaluate the perceptual quality of our method with four state-of-the-art methods: FeMaSR \cite{femasr:chen2022real}, LDL \cite{ldl:liang2022details}, SPSR \cite{spsr:ma2021structure}, and ESRGAN \cite{esrgan}. We invited 46 participants, ranking the results of each method on a scale from 1 (worst) to 5 (best) for 25 groups of images generated by the methods above. We collected a total of 1140 votes from the participants and analyzed the preferences of each method based on the percentage and number of votes for being in the Top 3 and Top 1 positions, and the average rank of each method. As illustrated in Fig.~\ref{fig:user_study}, the results demonstrate that our method outperforms the existing state-of-the-art methods, receiving more than half of the votes for being the best method and one-third of the votes for being in the top three methods. Our method also achieves superior perceptual quality and visual fidelity, as shown by qualitative and quantitative comparisons.


\section{Conclusion}\label{sec:conclu}
In this study, we have identified and defined the \emph{Center-oriented Optimization (COO) problem}, which is a critical obstacle in enhancing the perceptual quality of PSNR-oriented models. This discovery constitutes an important contribution to the theory of perceptual-driven super-resolution and opens up new avenues for the interpretation of existing methods. Our analysis suggests that many commonly used techniques, including perceptual loss, model optimization, and GAN-based methods, implicitly mitigate the COO problem by reducing the data entropy.
To address this problem, we have proposed a novel technique called \emph{Detail Enhanced Contrastive Loss (DECLoss)}. DECLoss utilizes the clustering property of contrastive learning to directly minimize the entropy of the potential HR distribution, thereby optimizing the COO problem. Our experimental results demonstrate the effectiveness and superiority of DECLoss on a range of SR benchmarks when combined with RaGAN. Specifically, DECLoss outperforms the state-of-the-art methods in both subjective metric LPIPS and objective metric PSNR. Notably, PSNR-oriented models can also enhance their perceptual quality by simply adding DECLoss. 
We plan to continue our research on this problem and refine our theory. Our future work will focus on exploring new methods to alleviate the COO problem and further improve the performance of perceptual-driven super-resolution models. Overall, our findings provide a deeper understanding of the COO problem and offer a promising approach for enhancing the perceptual quality of SR models.

\vspace{2em}


\appendices
\section{Comparisons of other Low-level Computer Vision tasks}

\begin{figure*}[!t]
\centering
\includegraphics[width=\linewidth]{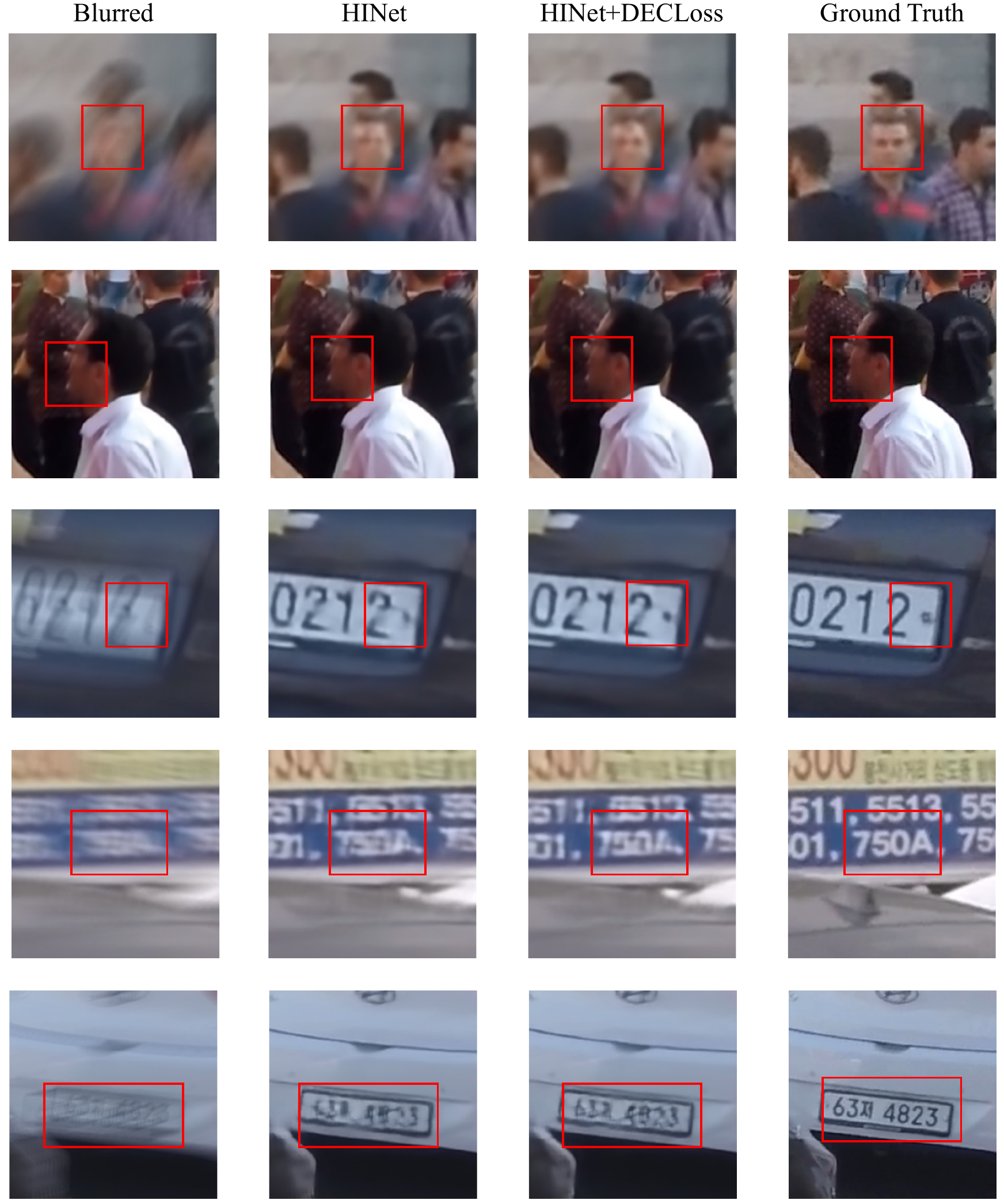}
\caption{Comparing with HINet on image deblurring (zoom in for the best view)} 
\label{fig:deblur}
\end{figure*}

\begin{figure*}[!t]
\centering
\includegraphics[width=\linewidth]{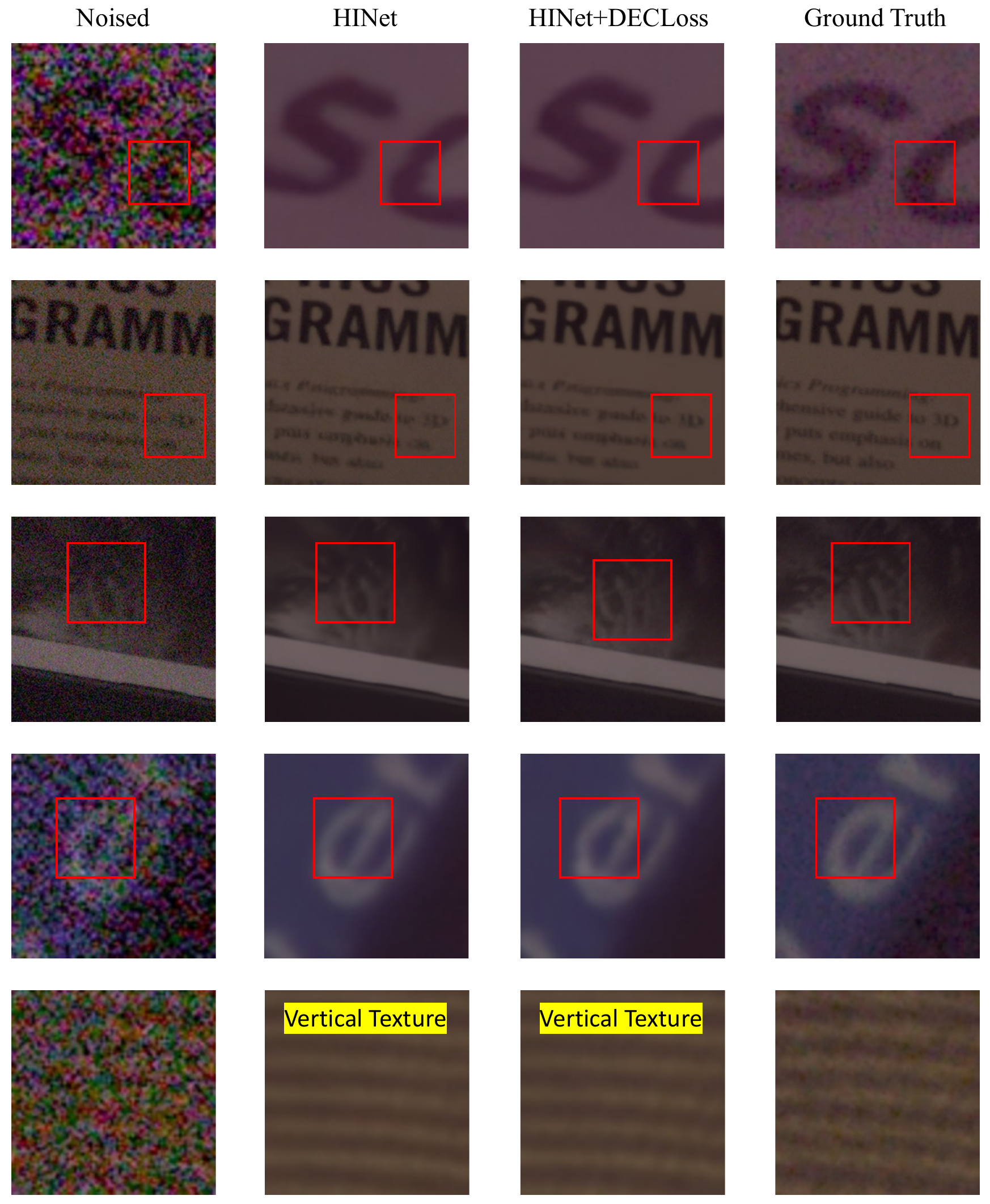}
\caption{Comparing with HINet on image denoising (zoom in for the best view)} 
\label{fig:denoise}
\end{figure*}

We have applied our proposed DECLoss to image deblurring and denoising tasks, using the 60G FLOPs version of HINet \cite{chen2021hinet} as the base model, which achieved the top performance in the 2021 NTIRE Image Deblurring Challenge. Figures~\ref{fig:deblur} and \ref{fig:denoise} demonstrate that DECLoss can produce more realistic textures than HINet across different categories, particularly for numbers and text. 
It is worth noting that the information loss, or loss of information entropy, in image denoising and deblurring tasks is smaller than that in image super-resolution. For instance, the image loss reaches 15/16 for 4x downsampling and 63/64 for 8x downsampling, which means the signal-to-noise ratio of downsampling is much smaller than noise or blur. As a result, the strength of the COO problem on these tasks is not as significant as image super-resolution. Therefore, explicit methods for optimizing the COO problem, such as DECLoss, have relatively limited potential for improving image quality.


\section{High-frequency Enhancement}
\label{sec:high-frequency}
We propose an alternative approach to enhance the high-frequency details of PSNR-oriented images by manipulating their Fourier coefficients. As depicted in Figure~\ref{fig:decloss_overview}(b), we employ the Discrete Fourier Transform (DFT), represented by $F$, to map both the super-resolved images and their corresponding ground truth, denoted as $Y$, to the Fourier domain $Y_f$:
\begin{equation}\label{eq:dft}
   Y_f = F^{(H)}YF^{(W)},
\end{equation}
where $H$ and $W$ are the height and width of an image $Y$, respectively. And each component of $F^{(H)}$ is defined by
\begin{equation}
   F_{jk}^{(N)} = \exp{(-2 \pi i j k / N)},\ i= \sqrt{-1}.
\end{equation}
We then multiply the inverse Gaussian kernel vector $K$ with the Fourier matrix to obtain
\begin{equation}\label{eq:gaussian_2}
   Y_f^\prime = \left(K^{(H)}\right)^T K^{(W)} \cdot Y_f,
\end{equation}
where each component of $K^{(H)}$ is defined by
\begin{equation}\label{eq:gaussian}
   K_i^{(N)} = \alpha \times \exp{\left(\frac{-(i-(N-1)/2)^2}{2 \times \mu ^2}\right)},
\end{equation}
where $\alpha$ and $\mu$ are control variables. Finally, we use the Inverse Discrete Fourier Transform (IDFT), denoted as $F^*$, to map the Fourier matrix $Y_f^*$ back to the image domain $Y^*$
\begin{equation}\label{eq:idft}
   Y^* = \frac{1}{HW}\left(F^{(H)}\right)^*Y_f\left(F^{(W)}\right)^*,
\end{equation}
where $F^{(N)}$ is defined as
\begin{equation}
   \left(F^{(N)}\right)^* = \left(\text{real}\left(F^{(N)}\right)-\text{imag}\left(F^{(N)}\right)\right)^T.
\end{equation}
Here, real$(\cdot)$ and imag$(\cdot)$ are the real and imaginary parts of a complex number, respectively. Note that, the inverse Gaussian kernel provides a smoothing filter for the regions of different frequencies, suppressing the low frequencies and enhancing the high frequencies, respectively.

\section{Selection of Data Distribution}
\label{sec:distribution_selection}

In our paper, we assume that all potential HR patches follow an unimodal Gaussian distribution. A multimodal distribution might seem more reasonable intuitively, but it poses practical challenges for calculating and relating it to our theory. For example, if we use a Gaussian Mixture Model (GMM) to describe the data, the probability of GMM $p_G$ can be written as
\begin{align}
    p_G(x|\theta) = \sum_{i=1}^{N}\alpha_i \phi(x|\theta_i),
\end{align}
where $\phi(x|\theta_i)$ denotes $i$-th unimodal Gaussian distribution
\begin{align}
    \phi(x|\mu, \Sigma) = \frac{1}{(2\pi)^n|\Sigma|^{\frac{1}{2}}}\exp{-\frac{(x-\mu)^T\Sigma^{-1}(x-\mu)}{2}}
\end{align}
and $\alpha_i$ is the corresponding factor. The information entropy of $p_G$ can be written as
\begin{align}
    H(x) &= - \int p_G(x|\theta)\ \log p_G(x|\theta)\ \text{dx} \\
    &= - \int \sum_{i=1}^{N}\alpha_{i}\ \phi(x|\theta_{i})\ \log \sum_{j=1}^{N}\alpha_{j}\ \phi(x|\theta_j) \ \text{dx} .
\end{align}

However, this expression cannot be computed in closed form due to the logarithm of a sum of exponential functions \cite{2022variational}. This is a well-known problem in machine learning, and many approaches have been proposed to address it by using Monte Carlo sampling \cite{moss2020cross, 2008on-entropy}, simplifying it as an unimodal Gaussian distribution \cite{2022variational}, or deriving upper and lower bounds to estimate it \cite{2008on-entropy, gershman2012nonparametric, kolchinsky2017estimating}. Therefore, it is very difficult to use GMM as the data distribution in our theory.

Moreover, based on the law of large numbers, it is reasonable to approximate the distribution as unimodal. This simplification enables us to efficiently capture the underlying structure of the data.

In summary, our method's choice to model HR patch distribution as an unimodal Gaussian distribution is rooted in the large amount of data and the statistical principles of large numbers. Using a multimodal distribution is theoretically complex and hard to implement in our proposed theory.

\section{Details of the Toy Example}
\label{app:toy_detail}

In this section, we provide the details of the toy example as illustrated in Fig.~4.
Suppose we have a binary classification problem where the input is a two-dimensional vector $x = (x_1, x_2)$ and the output is a binary label $y \in \{0, 1\}$. The data points are generated from two Gaussian distributions with different means and covariances,
\begin{equation}
    p(x|y=0) = N(\mu_0,\Sigma_0),~p(x|y=1)=N(\mu_1, \Sigma_1).
\end{equation}

We train a multi-layer perceptron (MLP) model to classify the data points. The MLP has one hidden layer with $n$ units and a sigmoid activation function, and one output layer with a single unit and a sigmoid activation function. 

As you can see from the output, increasing the number of parameters of the MLP model improves its discriminative power by making the decision boundary more flexible and adaptive to the data distribution. The MLP with 5 parameters can only draw a straight line as the decision boundary, which fails to separate the two classes well. The MLP with 9 parameters can draw a curved line as the decision boundary, which improves the classification performance slightly. The MLP with 33 parameters can draw a more complex curve as the decision boundary, which achieves a better separation of the two classes.


%




\section*{Acknowledgments}

This work was supported by the National Key R\&D Program of China (No.2022ZD0118202), the National Science Fund for Distinguished Young Scholars (No.62025603), the National Natural Science Foundation of China (No. U21B2037, No. U22B2051, No. 62176222, No. 62176223, No. 62176226, No. 62072386, No. 62072387, No. 62072389, No. 62002305 and No. 62272401), and the Natural Science Foundation of Fujian Province of China (No.2021J01002, No.2022J06001).





\bibliographystyle{IEEEtran}
\bibliography{main}

\newpage

\textbf{           }

\newpage

\textbf{      }

\newpage

\textbf{      }

\newpage

\textbf{      }

\newpage

%



%
\begin{IEEEbiography}[{\includegraphics[width=1in,clip,keepaspectratio]{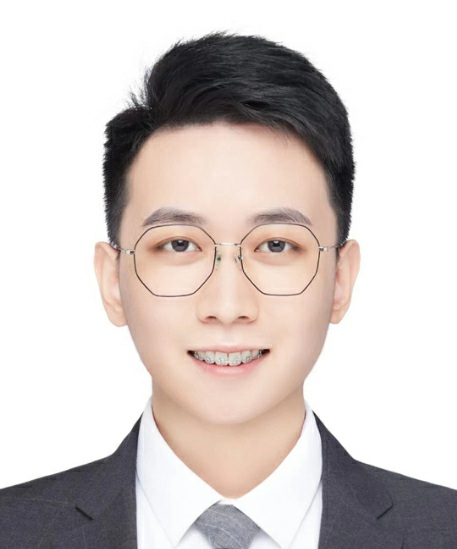}}]{Tianshuo Xu} received the M.S. and B.Sc. degree in intelligence science and technology from the School of Informatics, Xiamen University, Xiamen, China, in 2023 and 2020, respectively. He is currently working toward the Ph.D. degree in Artificial Intelligence at the Hong Kong University of Science and Technology (Guangzhou), China. His research interests include computer vision and machine learning.
\end{IEEEbiography}

\begin{IEEEbiography}[{\includegraphics[width=1in,clip,keepaspectratio]{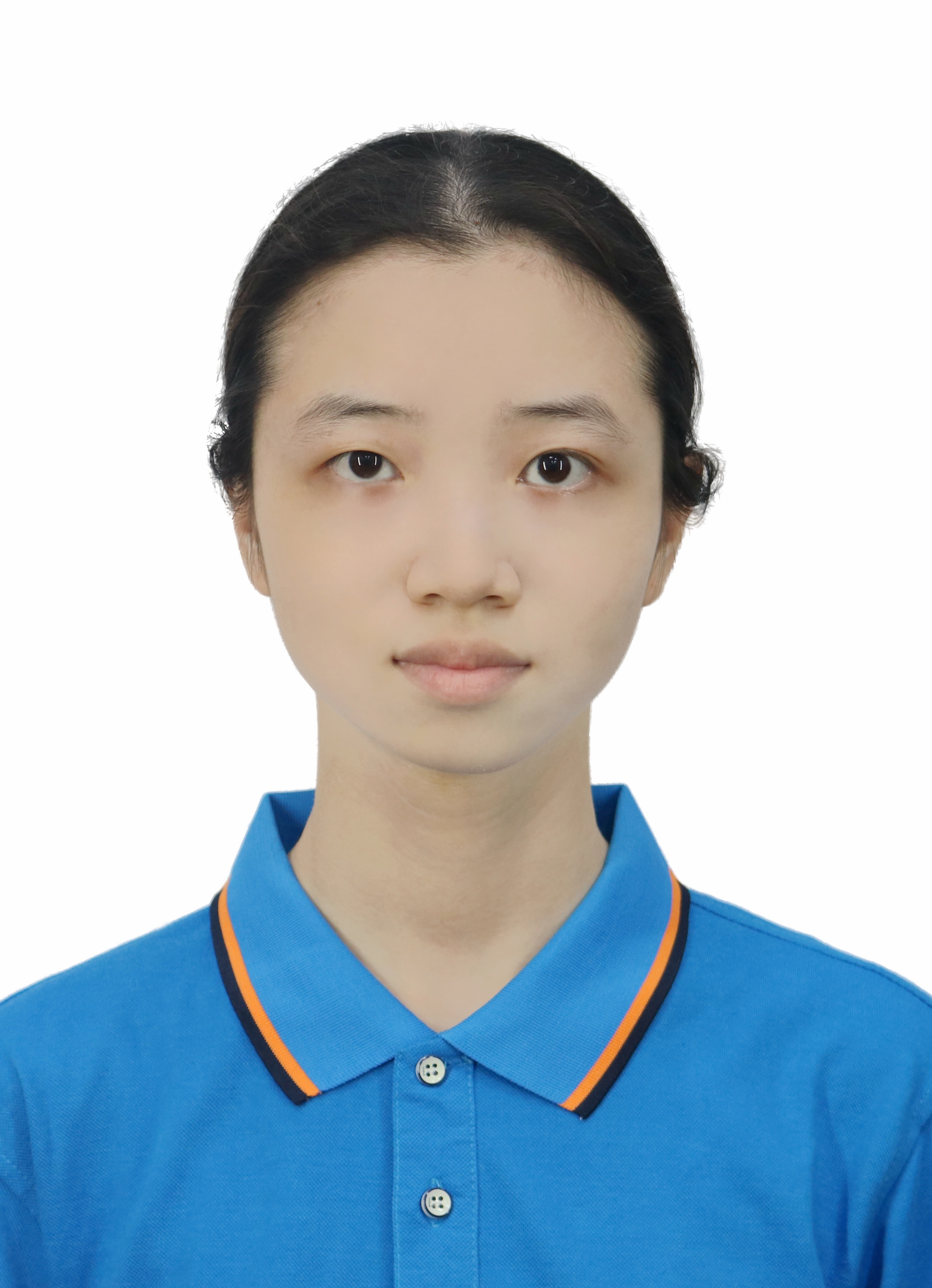}}]{Lijiang Li} received the B.Sc. degree in intelligence science and technology from the School of Informatics, Xiamen University, Xiamen, China, in 2022. She is currently working toward the M.S. degree in intelligence science and technology from the School of Informatics, Xiamen University, China. Her research interests include computer vision and machine learning.
\end{IEEEbiography}

\begin{IEEEbiography}[{\includegraphics[width=1in,clip,keepaspectratio]{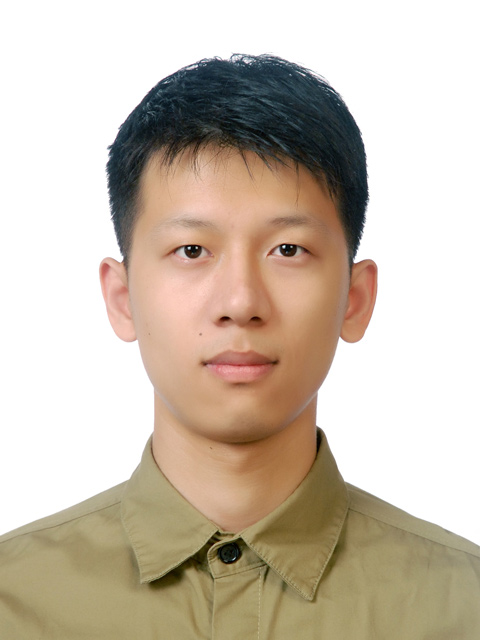}}]{Peng Mi} received the B.Sc. degree in Electronic Information Engineering, from Xidian University, Xi'an, China in 2020. He is currently working toward the M.S. degree in artificial intelligence from the School of Informatics, Xiamen University, China. His research interests include computer vision and machine learning.
\end{IEEEbiography}

\begin{IEEEbiography}[{\includegraphics[width=1in,clip,keepaspectratio]{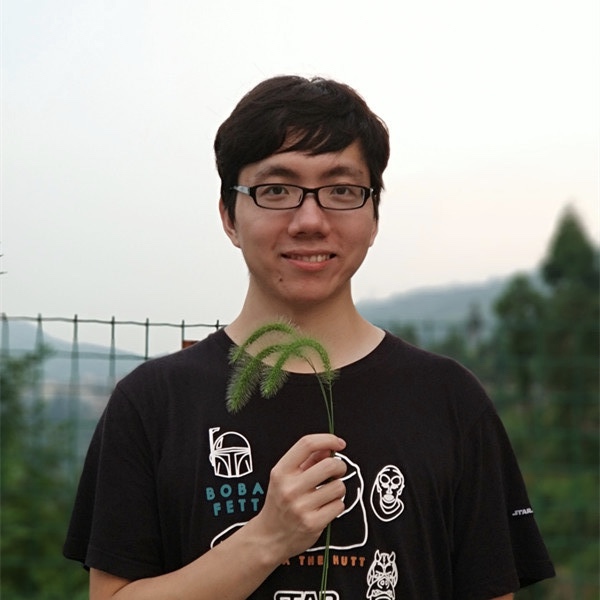}}]{Xiawu Zheng} received the MS degree in computer science from the School of Information Science and Engineering, Xiamen University, Xiamen, China, in 2018. He is currently working toward the Ph.D. degree from the School of Information Science and Engineering, Xiamen University, China. His research interests include computer vision and machine learning. He was involved in automatic machine learning.
\end{IEEEbiography}

\begin{IEEEbiography}[{\includegraphics[width=1in,clip,keepaspectratio]{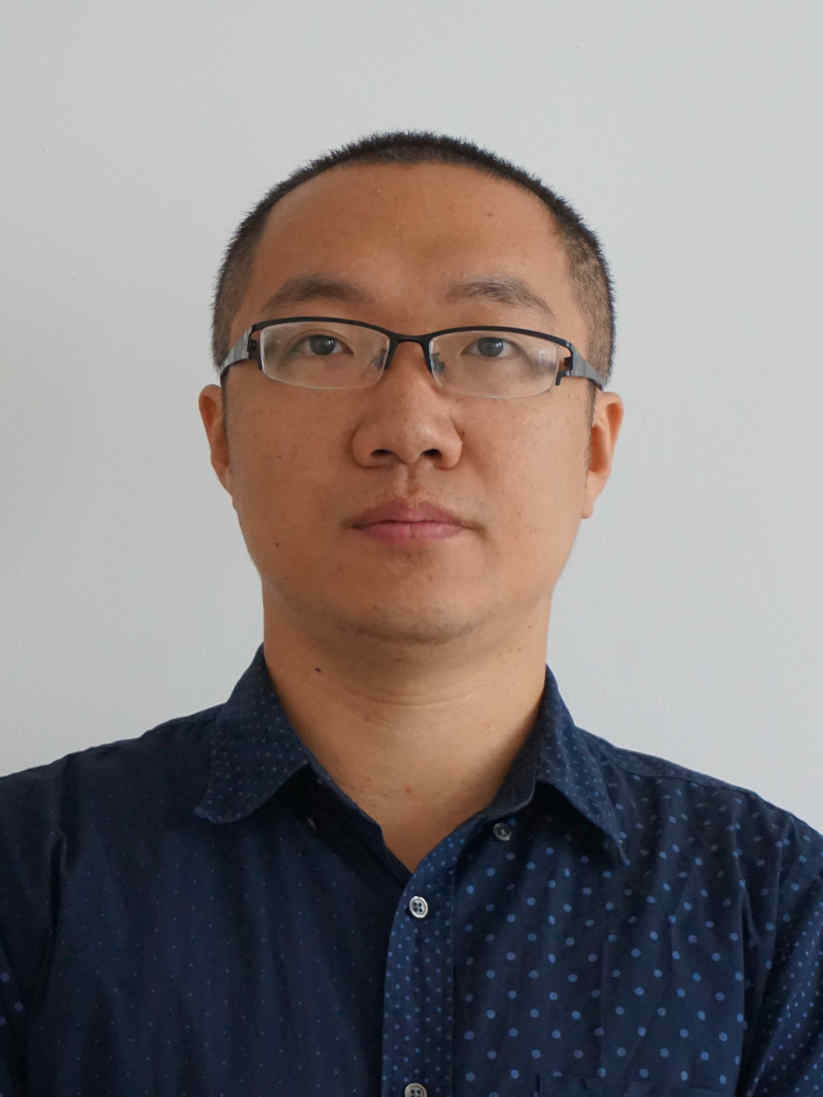}}]{Fei Chao} (Member, IEEE)
received the B.Sc. degree in Mechanical Engineering from Fuzhou University, China, and the M.Sc. Degree with distinction in Computer Science from the University of Wales, Aberystwyth, U.K., in 2004 and 2005, respectively, and the Ph.D. degree in robotics from the Aberystwyth University, Wales, U.K. in 2009. He is currently an Associate Professor with the School of Informatics, Xiamen University, China; and he is an adjunct research fellow with the Department of Computer Science, Aberystwyth University. Dr Chao has published more than 100 peer-reviewed journal and conference papers. His research interests include machine learning algorithms and network architecture search.
\end{IEEEbiography}

\begin{IEEEbiography}[{\includegraphics[width=1in,clip,keepaspectratio]{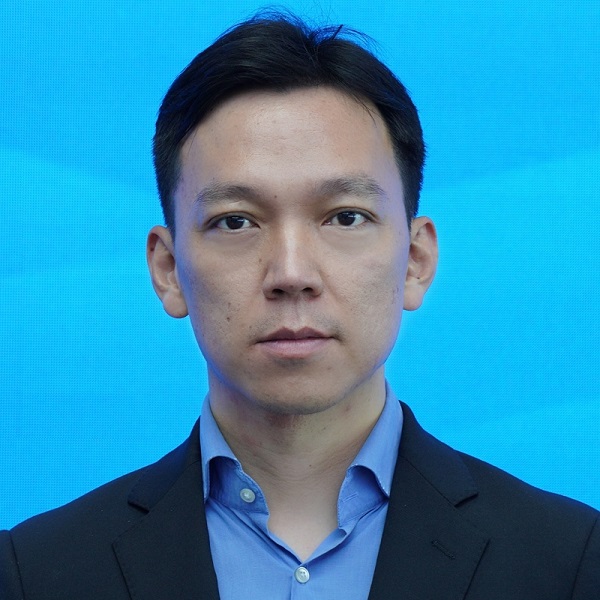}}]{Rongrong Ji} (Senior Member, IEEE)
 is currently a Nanqiang distinguished professor at Xiamen University, China, the deputy director of the Office of Science and Technology at Xiamen University, China, and the director of Media Analytics and Computing Lab. He was awarded as the National Science Foundation for Excellent Young Scholars (2014), the National Ten Thousand Plan for Young Top Talents (2017), and the National Science Foundation for Distinguished Young Scholars (2020). His research interests include computer vision, multimedia analysis, and machine learning. He has published more than 50 papers in ACM/IEEE Transactions, including the IEEE Transactions on Pattern Analysis and Machine Intelligence and the International Journal of Computer Vision, and more than 100 full papers on top-tier conferences, such as CVPR and NeurIPS. His publications have got more than 10K citations in Google Scholar. He was the recipient of the Best Paper Award of ACM Multimedia 2011. He has served as area chair in top-tier conferences such as CVPR and ACM Multimedia. He is also an advisory member for Artificial Intelligence Construction in the Electronic Information Education Committee of the National Ministry of Education.
\end{IEEEbiography}

\begin{IEEEbiography}[{\includegraphics[width=1in,clip,keepaspectratio]{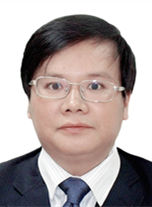}}]{Yonghong Tian} (Fellow, IEEE) is currently the Boya distinguished professor of the School of Electronics Engineering and Computer Science, Peking University, China, and also the deputy director of the Peng Cheng Laboratory, Artificial Intelligence Research Center, Shenzhen, China. His research interests include computer vision, multimedia big data, and brain-inspired computation. He is the author or co-author of more than 180 technical articles in refereed journals, such as the IEEE Transactions on Pattern Analysis and Machine Intelligence, the IEEE Transactions on Neural Networks and Learning Systems, the IEEE Transactions on Image Processing, the IEEE Transactions on Multimedia, the IEEE Transactions on Circuits and Systems for Video Technology, the IEEE Transactions on Knowledge and Data Engineering, the IEEE Transactions on Parallel and Distributed Systems and the ACM Computing Surveys, the ACM Transactions on Information Systems, the ACM Transactions on Multimedia Computing, Communications, and Applications and conferences, such as the NeurIPS/CVPR/ICCV/AAAI/ACMMM/WWW. He is a senior member of the CIE and CCF and a member of the ACM. He has been the steering member of the IEEE ICME since 2018 and the IEEE International Conference on Multimedia Big Data (BigMM) since 2015, and a TPC member of more than ten conferences, such as CVPR, ICCV, ACM KDD, AAAI, ACM MM, and ECCV. He was a recipient of the Chinese National Science Foundation for Distinguished Young Scholars, in 2018, two National Science and Technology Awards, three Ministerial-Level Awards in China, the 2015 EURASIP Best Paper Award for the Journal on Image and Video Processing, and the Best Paper Award of the IEEE BigMM 2018. He was/is an associate editor of the IEEE Transactions on Circuits and Systems for Video Technology since January 2018, the IEEE Transactions on Multimedia from August 2014 to August 2018, the IEEE Multimedia Magazine since January 2018, and the IEEE Access since 2017. He also Co-Initiated the BigMM. He has served as the TPC co-chair for BigMM 2015 and the Technical Program co-chair for the IEEE ICME 2015, the IEEE ISM 2015, and the IEEE MIPR 2018/2019, and the general co-chair for the IEEE MIPR 2020.
\end{IEEEbiography}

\begin{IEEEbiography}[{\includegraphics[width=1in,clip,keepaspectratio]{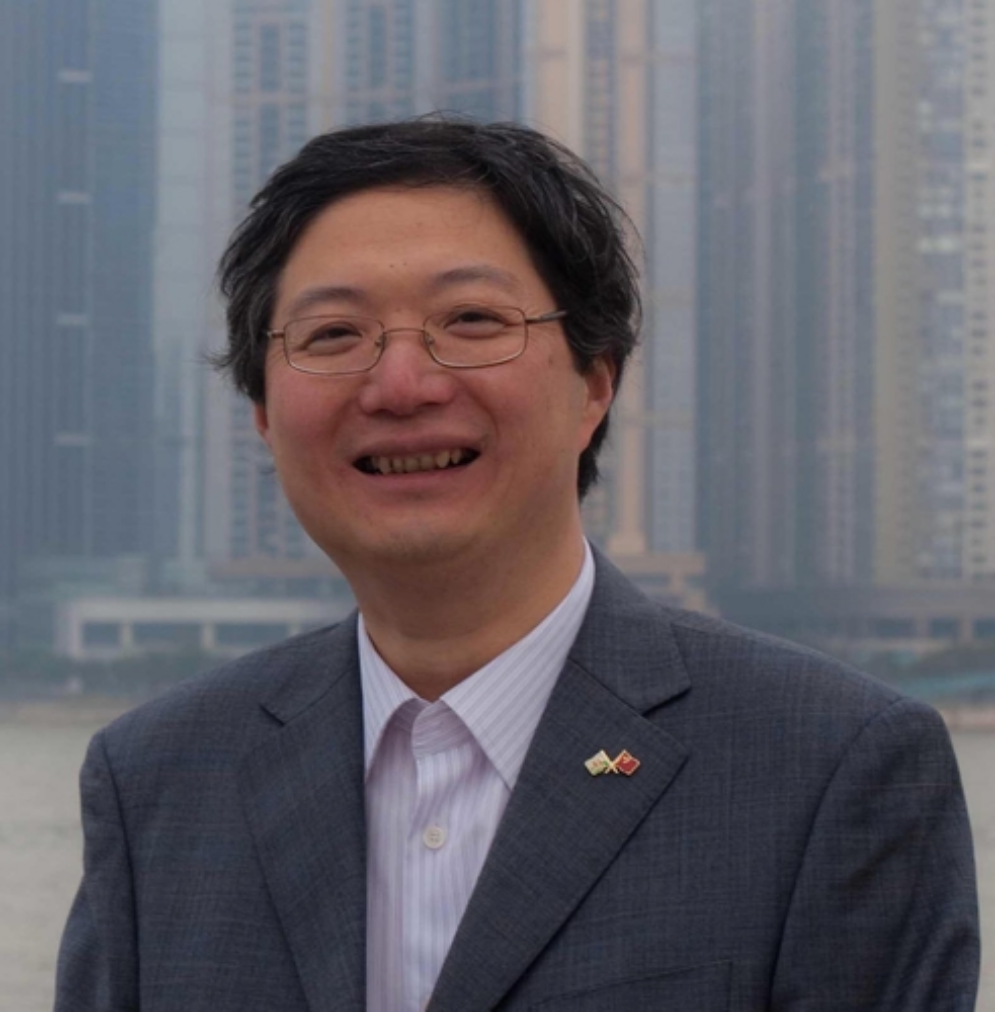}}]{Qiang Shen} is a Fellow of the Royal Academy of Engineer. He received the Ph.D. in Computing and Electrical Engineering (1990) from Heriot-Watt University, Edinburgh, U.K., and the D.Sc. in Computational Intelligence (2013) from Aberystwyth University, Aberystwyth, U.K. He holds the Established Chair in Computer Science and is the Pro Vice-Chancellor: Faculty of Business and Physical Sciences, Aberystwyth University. His research interests include computational intelligence and its application in robotics. He has authored three research monographs and over 470 peer-reviewed papers. Dr Shen is the recipient of the 2024 IEEE CIS Fuzzy Systems Pioneer Award.
\end{IEEEbiography}







\end{document}